\documentclass[fleqn,usenatbib]{mnras}

\usepackage{color}
\usepackage{graphicx}
\usepackage{amssymb}
\usepackage{amsmath}
\usepackage{bm}
\usepackage{bbold}
\usepackage{xparse}
\usepackage{xspace}
\xspaceaddexceptions{]\}}

\newcommand{\onefig}[3]{%
  \begin{figure}%
    \centerline{\resizebox{\hsize}{!}{\includegraphics*{#1}}}%
    \caption{#3}\label{#2}%
  \end{figure}%
}
\newcommand{\twofig}[4]{%
  \begin{figure*}%
    \centerline{\resizebox{\hsize}{!}{\includegraphics*{#1} \,%
        \includegraphics*{#2}}}%
    \caption{#4}\label{#3}%
  \end{figure*}%
}
\newcommand{\sect}[1]{Sect.~\ref{#1}\xspace}
\newcommand{\sects}[1]{Sects.~\ref{#1}\xspace}
\newcommand{\app}[1]{Appendix~\ref{#1}\xspace}
\newcommand{\fig}[1]{Fig.~\ref{#1}\xspace}
\newcommand{\eq}[1]{Eq.~(\ref{#1})\xspace}
\DeclareDocumentCommand{\eqs}{m m m o o}{%
  \IfNoValueTF {#4} {%
    Eqs.~(\ref{#1}){\xspace #2} (\ref{#3})%
  }{%
    Eqs.~(\ref{#1}){\xspace #2} (\ref{#3}){\xspace #4} (\ref{#5})%
  }%
}

\newcommand{\pcl}{pseudo-$C_{\ell}$\xspace}
\newcommand{\lmax}{\ensuremath{\ell_{\mathrm{max}}}\xspace}

\newcommand{\npix}{\ensuremath{n_{\mathrm{pix}}}\xspace}
\newcommand{\ntemp}{\ensuremath{n_{\mathrm{temp}}}\xspace}
\newcommand{\fsky}{\ensuremath{f_{\mathrm{sky}}}\xspace}
\newcommand\order[1]{${{\cal O}\! \left( #1 \right)}$}
\newcommand{\one}{\mathbb{1}}
\newcommand{\data}{\ensuremath{d}\xspace}
\newcommand{\tdata}{\ensuremath{\widetilde{\data}}\xspace}
\newcommand{\signal}{\ensuremath{s}\xspace}
\newcommand{\template}{\ensuremath{f}\xspace}
\newcommand{\templatev}{\ensuremath{\bm{f}}\xspace}
\newcommand{\mask}{w}
\newcommand{\fv}{\bm{F}}
\newcommand{\wv}{\bm{W}}
\newcommand{\uv}{\bm{U}}
\newcommand{\cv}{\bm{C}}
\newcommand{\tcv}{\bm{\widetilde{C}}}
\DeclareDocumentCommand{\alm}{O{} O{} O{a}}{\ensuremath{#3_{\ell{#1} m{#2}}}\xspace}
\newcommand{\eps}{\ensuremath{\epsilon}\xspace}
\newcommand{\eeps}{\hat{\epsilon}}
\newcommand{\cl}[2][]{C_{\ell{#1}}^{\, \mathrm{#2}}}
\newcommand{\clh}[2][]{\widehat{C}_{\ell{#1}}^{\, \mathrm{#2}}}
\newcommand{\cls}{\cl{\signal}}
\newcommand{\clt}{\clh{\template}}
\newcommand{\bn}{\bm{n}}
\newcommand{\dn}{\mathrm{d} \bn}
\newcommand{\llpo}[1][]{\ensuremath{(2{\ell{#1}} + 1)}\xspace}
\newcommand{\wigner}[6]{\ensuremath{\begin{pmatrix} #1 & #2 & #3 \\ #4 & #5 & #6 \end{pmatrix}}}

\newcommand{\equ}[2][]{\begin{equation}\label{#1}#2\end{equation}}
\newcommand{\eqa}[2][]{\begin{align}\label{#1}#2\end{align}}
\newcommand{\eqm}[2][]{\begin{multline}\label{#1}#2\end{multline}}
\newcommand{\nn}{\nonumber\\}

\newenvironment{referee}{\bf}{}
\newcommand{\bref}{\begin{referee}}
\newcommand{\eref}{\end{referee}}

\title[Pseudo-$C_{\ell}$ estimation with mode projection]{Unbiased \pcl power spectrum estimation with mode projection}

\author[Elsner et al.]{Franz Elsner,$^{1, 2}$\thanks{E-mail: f.elsner@ucl.ac.uk}
  Boris Leistedt,$^{1, 3}$
  and
  Hiranya V.~Peiris,$^{1, 4}$
\\
$^1$Department of Physics and Astronomy, University College London,
London WC1E 6BT, U.K.\\
$^2$Max--Planck--Institut f\"ur Astrophysik,
Karl--Schwarzschild--Stra\ss e 1, D--85748 Garching, Germany\\
$^3$Center for Cosmology and Particle Physics, Department of Physics,
New York University, New York, NY 10003, USA\\
$^4$The Oskar Klein Centre for Cosmoparticle Physics, Stockholm
University, 10691 Stockholm, Sweden
}

\date{Accepted \dots. Received \dots; in original form \dots}

\pubyear{2016}

\begin{document}
\label{firstpage}
\pagerange{\pageref{firstpage}--\pageref{lastpage}}
\maketitle

\begin{abstract}
  With the steadily improving sensitivity afforded by current and
  future galaxy surveys, a robust extraction of two-point correlation
  function measurements may become increasingly hampered by the
  presence of astrophysical foregrounds or observational
  systematics. The concept of mode projection has been introduced as a
  means to remove contaminants for which it is possible to construct a
  spatial map reflecting the expected signal contribution. Owing to
  its computational efficiency compared to minimum-variance methods,
  the sub-optimal \pcl (PCL) power spectrum estimator is a popular
  tool for the analysis of high-resolution data sets. Here, we
  integrate mode projection into the framework of PCL power spectrum
  estimation. In contrast to results obtained with optimal estimators,
  we show that the uncorrected projection of template maps leads to
  biased power spectra. Based on analytical calculations, we find
  exact closed-form expressions for the expectation value of the bias
  and demonstrate that they can be recast in a form that allows a
  numerically efficient evaluation, preserving the favorable
  \order{\lmax^3} time complexity of PCL estimator algorithms. Using
  simulated data sets, we assess the scaling of the bias with various
  analysis parameters and demonstrate that it can be reliably removed.
  We conclude that in combination with mode projection, PCL estimators
  allow for a fast and robust computation of power spectra in the
  presence of systematic effects -- properties in high demand for the
  analysis of ongoing and future large scale structure surveys.
\end{abstract}


\section{Introduction}
\label{sec:intro}

In modern cosmology, measurements of the power spectrum (or its
real-space counterpart, the angular correlation function) have proven
a powerful summary statistic and are widely used to confront
theoretical models with observational data, e.g.,
\citet{1992ApJ...396L...1S, 1994Natur.367..333H, 1995ApJ...443L..57G,
  1997ApJ...474...47N, 2000ApJ...545L...5H, 2002ApJ...568...38H,
  2002Natur.420..772K, 2003ApJS..148..135H, 2010ApJ...722.1148F,
  2010ApJ...719.1045L, 2014A&A...571A..15P, 2014ApJ...794..171T,
  2015PhRvL.114j1301B} for an arbitrary selection of measurements of
the cosmic microwave background radiation (CMB) two-point correlation
function, or, e.g., \citet{1969PASJ...21..221T, 1996MNRAS.283..709H,
  2001MNRAS.328...64N, 2002MNRAS.329L..37B, 2002ApJ...571..172Z,
  2004ApJ...606..702T, 2005MNRAS.356..415C, 2005ApJ...633..560E,
  2008ApJ...672..153C, 2010MNRAS.404...60R, 2011MNRAS.416.3017B,
  2014MNRAS.438..825K, 2016MNRAS.455.4301C} for constraints on galaxy
clustering.

A decrease in statistical errors resulting from the increasing
coverage or sensitivity of ongoing and future experiments will impose
stricter limits on the level of contamination of the targeted
cosmological signal by secondary sources. Such contaminants may be of
astrophysical origin (e.g., foreground emission or dust extinction,
e.g., \citealt{2005ApJ...619..147M}) or the result of complications
associated with the data collection and processing procedure (for
example, survey depth fluctuations, varying seeing conditions, image
calibration uncertainties, \citealt{2013MNRAS.432.2945H,
  2016ApJ...829...50A}). To aid assessment of the possible impact of
systematic effects that may have altered the observed signal, it has
become standard for galaxy surveys to compile libraries of template
maps that describe the spatial variation of survey properties
\citep{2002ApJ...579...48S, 2011MNRAS.417.1350R, 2012MNRAS.424..564R,
  2014MNRAS.444....2L, 2016ApJS..226...24L, 2016arXiv160703145R}.
Several approaches have been proposed that make use of these maps to
correct measurements of the two-point statistics for systematic
effects (\citealt{1992ApJ...398..169R, 2012ApJ...761...14H,
  2014MNRAS.444....2L}, see \citealt{2016MNRAS.456.2095E} for a
comparison). In \citet{2016MNRAS.463..467K}, the authors derive a
template cleaning procedure for the popular FKP estimator
\citep{1994ApJ...426...23F}.

In the following, we focus on the mode projection procedure of
\citet{1992ApJ...398..169R}. Attributing infinite variance to modes
described by a set of templates, specific signal patterns can be
excluded from the analysis and the computed result hence becomes more
robust with respect to systematics captured by them \citep[see,
  e.g.,][for applications]{1998ApJ...499..555T, 2004PhRvD..69l3003S,
  2009JCAP...09..006S, 2013A&A...549A.111E,
  2013MNRAS.435.1857L}. Unfortunately, mode projection can only be
straightforwardly implemented in case the estimator makes use of
inverse variance-weighted data. Within the field of power spectrum
estimation, this is the case for the maximum likelihood estimator
\citep{1998PhRvD..57.2117B} and the optimal quadratic estimator
\citep{1997PhRvD..55.5895T}. Regrettably, both of them are very
expensive to evaluate numerically, usually prohibitively so for
state-of-the-art high-resolution data
\citep{1999PhRvD..59b7302B}. Conversely, the much faster \pcl (PCL)
estimator introduced by \citet{2002ApJ...567....2H} makes no attempt
at exact inverse variance-weighting, trading optimality for
computational speed, and can be applied in only \order{\lmax^3} time
to a data set band-limited at multipole moment \lmax. The purpose of
this paper is to demonstrate that the concept of mode projection can
be successfully integrated into the framework of PCL estimators,
combining the desirable properties of fast and robust power spectrum
estimation.

This article is organized as follows. In \sect{sec:theory}, we review
the concept of mode projection and discuss how it can be implemented
in PCL estimators. Then, we use numerical simulations to verify our
results and systematically study the impact of mode projection for
different analysis parameters (\sect{sec:verify}). We conclude by
summarizing our findings in \sect{sec:conclusions}.

\section{PCL mode projection}
\label{sec:theory}

We start this section by providing a detailed review of mode
projection \citep{1992ApJ...398..169R}. Straightforwardly integrated
into optimal power spectrum estimators, it was shown to lead to
unbiased results at the cost of an increase in the estimator variance
that is modest compared to other systematics mitigation schemes
\citep{2016MNRAS.456.2095E}.

We first consider a contaminant that can be described by a single
non-vanishing template \template and contributes with unknown scalar
amplitude \eps to the data vector \data,
\equ[eq:data_model_single_tpl]{
  \data = \signal + \eps \template \, .
}
Even if the simple linear model in \eq{eq:data_model_single_tpl} is
not fully appropriate, we can still use it as a first order
approximation of a Taylor expansion in \template for small values of
\eps. In the following, we assume the absence of correlations between
stochastic signal realizations \signal and the deterministic template
\template used in the projection in the ensemble average.

Then, our goal is to find a means to infer the power spectrum of the
targeted cosmological signal \signal,
\equ[eq:powers]{
  \clh{\signal} = \sum_{m} \frac{1}{2\ell + 1} \left|
  \alm[][][\signal] \right|^2 \, ,
}
where we have introduced the ``hat'' notation to specify an estimated
quantity for a specific realization of the analyzed field.

In case an analysis is based on inverse variance-weighted data only,
mode projection is implemented by modifying the data covariance matrix
$\cv$. A rank-one term, constructed from the template, is added with
variance $\sigma$. Afterwards, we take the limit to assign infinite
variance to this specific signal direction,
\equ[eq:mpdef]{
  \tcv = \lim_{\sigma \to \infty} \left( \cv + \sigma \template
  \template^{\dagger} \right) \, .
}
Then, any analysis making use of the data $\data$ in form of
\equ[eq:projection]{
  \tdata = \tcv^{-1} \data
}
will be insensitive to a contaminant described by the template.

Guided by \eqs{eq:mpdef}{and}{eq:projection}, we now implement mode
projection within the framework of \pcl power spectrum
estimation. Since PCL does not make use of inverse variance-weighted
maps, we apply the PCL estimator to a filtered version of the
data. The filter is linear and can be expressed in terms of a matrix,
\equ[eq:pclfilter]{
  \fv = \lim_{\sigma \to \infty} \left( \one + \sigma \template
  \template^{\dagger} \right)^{-1} \, ,
}
where $\one$ is the identity matrix. Making use of the
Sherman-Morrison formula, we can take the limit and find an
exact expression for the filter,
\equ[eq:pclfiltersingle]{
  \fv = \one - \frac{\template
    \template^{\dagger}}{\template^{\dagger} \template} \, .
}
We therefore derive for the preprocessed data vector $\tdata = \fv
\data$,
\equ[eq:filtered_data]{
  \tdata = \data - \frac{\template^{\dagger}
    \data}{\template^{\dagger} \template} \, \template \, .
}
From \eq{eq:filtered_data} the well known equivalence between mode
projection and a direct subtraction becomes apparent again
\citep{1992ApJ...398..169R}, i.e., the data are cleaned by removing a
template contribution with amplitude estimate $\eeps =
\left. \template^{\dagger} \data \middle/ \template^{\dagger}
\template \right.$.

As a side note, we mention that \eq{eq:filtered_data} represents the
simplest case where all modes are assigned equal weights in the
calculation of the cleaning coefficient $\eeps$. Relaxing this
assumption would require introducing a weight matrix $\wv$ such that
$\eeps = \left. \template^{\dagger} \wv \data \middle/
\template^{\dagger} \wv \template \right.$. For $\wv = \cv^{-1}$, we
then recover the maximum likelihood cleaning approach that is
implicitly used in optimal mode projection algorithms. Since it is
possible to construct the Cholesky decomposition $\wv = \uv^{\dagger}
\uv$ for any given positive-definite weight matrix, we can choose to
consider the prewhitened data vector $\data_\mathrm{w} = \uv \data$
instead, and absorb all remaining factors of $\uv$ by redefining
$\template_\mathrm{w} = \uv \template$, leading back to
\eq{eq:filtered_data}. We can therefore set the weight matrix to unity
in what follows.

As we will demonstrate below, even in the absence of any contaminant,
applying a power spectrum estimator to \tdata to measure the
statistical properties of \signal will in general lead to biased
results. We now derive analytical expressions for the expectation
value of the bias introduced by mode projection. We begin our
discussion by analyzing the simplest possible case, the projection of
a single template on the full sky, and then gradually generalize our
findings to take into account the effects of multiple templates and
limited sky coverage. Readers only interested in our main result may
skip the first paragraphs and continue with
\sect{sec:sub_cut_sky_multi}. In our calculation, we will assume that
the Fourier modes of the field analyzed are mutually uncorrelated to
sufficient precision in the ensemble average, i.e., $\langle
\alm[][][\signal] \ \alm[^{\prime}][^{\prime}][\signal]^{\ast} \rangle
\propto \delta_{\ell \ell^{\prime}} \delta_{m
  m^{\prime}}$.\footnote{The same assumption must be made in the
  derivation of the PCL estimator, \citet{2002ApJ...567....2H}.}

\subsection{Full sky analysis, single template}
\label{sec:sub_full_sky_single}

We start out considering a full sky analysis where a single template
has been projected out. Given the filtered data map
\eq{eq:filtered_data} as input, we derive the mean variance of its
spherical harmonic coefficients,
\eqm[eq:fullsky]{
  \langle \alm[][][\widetilde{\data}]
  \ \alm[][][\widetilde{\data}]^{\ast} \rangle
  = \langle \alm[][][\signal] \ \alm[][][\signal]^{\ast} \rangle -
  \frac{2}{\template^{\dagger} \template} \langle \alm[][][\signal]
  \ (\signal^{\dagger} \template) \alm[][][\template]^{\ast} \rangle \\
  + \frac{1}{\left( \template^{\dagger} \template \right)^2} \langle
  (\template^{\dagger} \signal) \alm[][][\template] \ (\signal^{\dagger}
  \template) \alm[][][\template]^{\ast} \rangle \, .
}
Denoting the power spectrum of the template realization used in the
projection as $\clt$, we obtain for the normalization factor in the
above expression
\equ[eq:ftf]{
  \template^{\dagger} \template = \sum_{\ell} \llpo \clt \, ,
}
a measure of the total variance of \template. Introducing further
$\cls = \langle \clh{\signal}\rangle$ as the ensemble averaged signal
power spectrum, we derive the following expectation values for the
multipole moments $(\ell, m)$,%
\equ[eq:ssff_full_exact]{
  \langle \alm[][][\signal] \ (\signal^{\dagger} \template)
  \alm[][][\template]^{\ast} \rangle = \cls \clt \, ,
}
\equ[eq:ssffff_full_exact]{
  \langle (\template^{\dagger} \signal) \alm[][][\template]
  \ (\signal^{\dagger} \template) \alm[][][\template]^{\ast} \rangle =
  \left( \sum_{\ell^{\prime}} \llpo[^{\prime}]
  \cl[^{\prime}]{\signal} \clh[^{\prime}]{\template} \right) \clt \, .
}
Projecting a single template on the full sky, the ensemble averaged
power spectrum of the filtered data set $\tdata$ becomes
\eqa{
  \langle \clh{\tdata} \rangle &= \sum_{m} \frac{1}{2\ell +
    1} \langle \alm[][][\tdata] \ \alm[][][\tdata]^{\ast} \rangle \nn
  &= \cls + b_{\ell} \, .
}
Since we want to use $\clh{\widetilde{\data}}$ as a proxy for the
signal power spectrum $\clh{\signal}$, we conclude that this estimate
is biased. The bias $b_{\ell}$ is given by
\equ[eq:fullsky_bias]{
b_{\ell} = - \frac{2 \cls \clt}{\sum_{\ell^{\prime}} \llpo[^{\prime}]
  \clh[^{\prime}]{\template}} + \frac{\left( \sum_{\ell^{\prime}}
  \llpo[^{\prime}] \cl[^{\prime}]{\signal} \clh[^{\prime}]{\template}
  \right) \clt}{\left( \sum_{\ell^{\prime}} \llpo[^{\prime}]
  \clh[^{\prime}]{\template} \right)^2} \, .
}

We therefore obtain a simple recipe to combine mode projection and PCL
power spectrum estimation. Instead of directly analyzing a given data
set, we first apply a filter function according to
\eq{eq:filtered_data}. After the power spectrum has been computed, the
result is then corrected by subtracting a bias term,
\equ[eq:debiasing]{
  \clh{\signal} = \clh{\tdata} - b_{\ell} \, ,
}
leading to clustering estimates of the signal
that are unbiased in the ensemble average and have been marginalized
over contaminants described by the template.

An additional complication in the evaluation of \eq{eq:debiasing}
arises from the fact that the bias term in itself is a function of the
signal power spectrum. In the full-sky case, it is still feasible to
compute $\clh{\signal}$ directly by finding the solution to the matrix
equation
\equ[eq:debiasing_matrix]{
  \clh{\signal} = \sum_{\ell^{\prime}} \left[ \left( \one + B
  \right)^{-1} \right]_{\ell \ell^{\prime}} \clh[^{\prime}]{\tdata} \, ,
}
where
\equ[eq:bias_matrix]{
  B_{\ell_1 \ell_2} = - \frac{2
    \clh[_1]{\template}}{\sum_{\ell^{\prime}} \llpo[^{\prime}]
    \clh[^{\prime}]{\template}} \cdot \delta_{\ell_1 \ell_2} +
  \frac{\llpo[_2] \clh[_2]{\template}
    \clh[_1]{\template}}{\left(\sum_{\ell^{\prime}} \llpo[^{\prime}]
    \clh[^{\prime}]{\template}\right)^2} \, .
}
It is interesting to note that even though we compute power spectra on
the full sky, $B$ will in general contain off-diagonal entries. We
conclude that applying the filter \eq{eq:filtered_data} can lead to
the coupling of previously uncorrelated Fourier modes. This behaviour
is in line with the interpretation that mode projection is equivalent
to masking (see \app{sec:mp_mask} for a detailed discussion).

We will later see that it is not always possible to find an explicit
expression for \eq{eq:bias_matrix}. In practice, it may therefore be
most viable to debias the result iteratively, or, assuming a prior
power spectrum for $\cls$.

\subsection{Full sky analysis, multiple templates}
\label{sec:sub_full_sky_multi}

To be able to handle multiple (not necessarily linearly independent)
templates requires a generalization of the filter matrix used to
prepare the data. For a data vector with \npix elements, we modify
\eq{eq:pclfiltersingle} to take a $\npix \times \ntemp$ object
$\templatev$ as input, containing a collection of \ntemp templates,
\equ[eq:pclfiltermulti]{
  \fv = \one - \templatev \left(\templatev^{\dagger} \templatev
  \right)^{-1} \templatev^{\dagger} \, ,
}
where the normalization factor now becomes a $\ntemp \times \ntemp$
matrix with entries computed from template auto- and cross-power
spectra,
\equ[eq:norm_factor]{
  \left(\templatev^{\dagger} \templatev \right)_{i j} = \sum_{\ell}
  \, \llpo \clh{\template^i \times \template^j} \, .
}
We propose to use the Moore-Penrose inverse for
$\left(\templatev^{\dagger} \templatev \right)^{-1}$ in case this
matrix is rank deficient.\footnote{One might encounter this situation,
  for example, in case there exists a $i \neq j$ for which
  $\template^i \propto \template^j$. If the pseudo inverse of
  $\left(\templatev^{\dagger} \templatev \right)$ is used for the
  inversion, such degeneracies are taken into account fully
  self-consistently by the algorithm. This property obviates the need
  to check a potentially large template library for linear
  dependencies.}

Projecting multiple templates on the full sky, \eq{eq:fullsky} now
takes the form
\eqm[eq:fullskymulti]{
  \langle \alm[][][\widetilde{\data}]
  \ \alm[][][\widetilde{\data}]^{\ast} \rangle
  = \langle \alm[][][\signal] \ \alm[][][\signal]^{\ast} \rangle - 2
  \sum_{i j} \left(\templatev^{\dagger} \templatev \right)^{-1}_{i j}
  \langle \alm[][][\signal] \ (\signal^{\dagger} \template^j)
  \alm[][][\template]^{i \, \ast} \rangle \\
  + \sum_{\substack{i j\\ h k}} \left(\templatev^{\dagger}
  \templatev \right)^{-1}_{i j} \left(\templatev^{\dagger} \templatev
  \right)^{-1}_{h k} \langle (\template^{j \, \dagger} \signal)
  \alm[][][\template]^{i} \ (\signal^{\dagger} \template^{k})
  \alm[][][\template]^{h \, \ast} \rangle \, ,
}
and we find for the bias
\eqm[eq:fullskymulti_bias]{
  b_{\ell} = -2 \sum_{i j} \left(\templatev^{\dagger} \templatev
  \right)^{-1}_{i j} \cls \clh{\template^j \times \template^i} +
  \sum_{\substack{i j\\ h k}} \left(\templatev^{\dagger} \templatev
  \right)^{-1}_{i j} \left(\templatev^{\dagger} \templatev
  \right)^{-1}_{h k} \\
  \times \left(\sum_{\ell^{\prime}} \llpo[^{\prime}]
  \cl[^{\prime}]{\signal} \clh[^{\prime}]{\template^j \times
    \template^k} \right) \clh{\template^i \times \template^h} \, ,
}
the generalization of \eq{eq:fullsky_bias} to multiple templates.
With this result, we can trivially provide an explicit expression for
the generalized bias matrix \eq{eq:bias_matrix} that can be used with
\eq{eq:debiasing_matrix} to obtain unbiased signal power spectrum
estimates,
\eqm{
  B_{\ell_1 \ell_2} = -2 \sum_{i j} \left(\templatev^{\dagger}
  \templatev \right)^{-1}_{i j} \clh[_1]{\template^j \times \template^i}
  \cdot \delta_{\ell_1 \ell_2} \\
  + \sum_{\substack{i j\\ h k}} \left(\templatev^{\dagger} \templatev
  \right)^{-1}_{i j} \left(\templatev^{\dagger} \templatev
  \right)^{-1}_{h k} \llpo[_2] \clh[_2]{\template^j \times
    \template^k} \clh[_1]{\template^i \times \template^h} \, .
}

\subsection{Cut-sky analysis, single template}
\label{sec:sub_cut_sky_single}

We now turn to the more realistic case of a cut-sky analysis. To allow
a transparent discussion of the problems associated with this
complication, we again start by first considering the projection of a
single template before generalizing our results later on.

Denoting \alm[][][a^{\mathrm{full}}] as the spherical harmonics of a
field on the full sky, a modified set of coefficients
\alm[][][a^{\mathrm{cut}}] is then obtained by multiplying its real
space representation with a non-negative mask $W$,
\eqa[eq:pcl_mode_coupling]{
  \alm[][][a^{\mathrm{cut}}] &= \sum_{\ell^{\prime} m^{\prime}}
  \alm[^{\prime}][^{\prime}][a^{\mathrm{full}}] \int \dn \,
  Y_{\ell^{\prime} m^{\prime}}(\bn) \, W(\bn) \, Y_{\ell
    m}^{\ast}(\bn) \nn
 &= \sum_{\ell^{\prime} m^{\prime}}
  \alm[^{\prime}][^{\prime}][a^{\mathrm{full}}] K_{\ell m \ell^{\prime}
    m^{\prime}} \, .
}
Here, the coupling kernels $K$ capture how the orthogonality relation
of the spherical harmonics is modified by the mask. We give their
exact definition in \app{sec:app_pcl} (\eq{eq:coupling_kernels}).

A \pcl power spectrum estimation algorithm then makes use of the
properties of the coupling kernels to obtain a simplified expression
that relates power spectra on the full sky to cut-sky spectra with
correct properties in the ensemble average,
\equ[eq:pcl_mask_convolution]{
  \langle \clh{full} \rangle = \sum_{\ell^{\prime}} M^{-1}_{\ell
    \ell^{\prime}} \langle \clh[^{\prime}]{cut} \rangle \, ,
}
where we have assumed that the inverse of the coupling matrix $M$
exists, a function of the mask power spectrum only (see
\eq{eq:coupling_matrix} for a formal definition).

Based on the framework developed for PCL estimators, it is now
possible to compute the mode projection bias (\eq{eq:fullsky_bias})
for limited sky coverage. In this case, data and template maps are
both multiplied with the mask prior to the analysis. We find the
expression of the normalization factor \eq{eq:ftf} to be unchanged,
although it is now calculated from cut-sky template pseudo-power
spectra that have not been corrected for the reduced sky
fraction. Computing the remaining terms, however, is more
complicated. We now obtain
(cf.\ \eqs{eq:ssff_full_exact}{and}{eq:ssffff_full_exact}),
\eqa[eq:ssff_cut_exact_alm]{
  \langle \alm[][][\signal] \ (\signal^{\dagger} \template)
  \alm[][][\template]^{\ast} \rangle &= \sum_{\substack{\ell_{1, 2, 3,
        4} \\ m_{1, 2, 3, 4}}} \! \cl[_2]{\signal}
  \ \alm[_1][_1][\template] \, \alm[][][\template]^{\ast}
  \ \alm[_3][_3][\mask] \, \alm[_4][_4][\mask]^{\ast} \nn
  & \times \sqrt{(2\ell + 1)(2\ell_1 + 1)(2\ell_3 + 1)(2\ell_4 + 1)} \nn
  & \times \frac{2\ell_2 + 1}{4 \pi}
  \wigner{\ell}{\ell_{2}}{\ell_{3}}{0}{0}{0}
  \wigner{\ell_{1}}{\ell_{2}}{\ell_{4}}{0}{0}{0} \nn
  & \times
  \wigner{\ell}{\ell_{2}}{\ell_{3}}{m}{-m_{2}}{m_{3}}
  \wigner{\ell_{1}}{\ell_{2}}{\ell_{4}}{m_{1}}{-m_{2}}{m_{4}} \, ,
}
where the last four objects are Wigner 3j symbols, and
\eqa[eq:ssffff_cut_exact_alm]{
  \langle (\template^{\dagger} \signal) \alm[][][\template]
  &\ (\signal^{\dagger} \template) \alm[][][\template]^{\ast} \rangle \nn
  &= \sum_{\substack{\ell_{1, 2, 3, 4, 5} \\ m_{1, 2, 3, 4, 5}}} \!
  \cl[_3]{\signal} \ \alm[][][\template] \, \alm[][][\template]^{\ast}
  \ \alm[_2][_2][\template] \, \alm[_1][_1][\template]^{\ast}
  \ \alm[_4][_4][\mask] \, \alm[_5][_5][\mask]^{\ast} \nn
  & \times \sqrt{(2\ell_1 + 1)(2\ell_2 + 1)(2\ell_4 + 1)(2\ell_5 + 1)} \nn
  & \times \frac{2\ell_3 + 1}{4 \pi}
  \wigner{\ell_{1}}{\ell_{3}}{\ell_{4}}{0}{0}{0}
  \wigner{\ell_{2}}{\ell_{3}}{\ell_{5}}{0}{0}{0} \nn
  & \times
  \wigner{\ell_{1}}{\ell_{3}}{\ell_{4}}{m_{1}}{-m_{3}}{m_{4}}
  \wigner{\ell_{2}}{\ell_{3}}{\ell_{5}}{m_{2}}{-m_{3}}{m_{5}} \nn
  &= \alm[][][\template] \, \alm[][][\template]^{\ast} \sum_{\ell_{1}
    m_{1}} \langle \alm[_1][_1][\signal] \ (\signal^{\dagger} \template)
  \alm[_1][_1][\template]^{\ast} \rangle \, .
}
While the above equations formally are the full solution to the
problem, we note that their brute force evaluation is in fact more
expensive than computing the optimal quadratic estimator with mode
projection, rendering the result useless for all practical purposes.

Luckily, we can substantially speed up the bias calculation in case of
limited sky coverage by leveraging the power of the convolution
theorem. Building on the properties of the Wigner 3j symbols
(\eq{eq:gaunt}), we use a mix of real- and spherical harmonic space
representations to transform \eq{eq:ssff_cut_exact_alm}, finding
\eqm[eq:ssff_cut_exact]{
  \langle \alm[][][\signal] \ (\signal^{\dagger} \template)
  \alm[][][\template]^{\ast} \rangle = (-1)^{m}
  \alm[][][\template]^{\ast} \int \dn \\
  \left( \sum_{\ell_{2} m_{2}} (-1)^{m_2} \cl[_2]{\signal} \! \left[
    \int \! \dn^{\prime} \, \template(\bn^{\prime}) \,
    \overline{W}(\bn^{\prime}) \, Y_{\ell_2 m_2}^{\ast}(\bn^{\prime})
    \right] \! Y_{\ell_2 m_2}(\bn) \! \right) \\
  \times \widetilde{W}(\bn) \, Y_{\ell m}^{\ast}(\bn) \, ,
}
where $\overline{W}$ and $\widetilde{W}$ are modified representations
of the mask in pixel space, computed from its spherical harmonic
coefficients, $\alm[][][\mask]$,
\eqa{
  \label{eq:mask_w_bar}
  \overline{W} &= \sum_{\ell m} \alm[][][\mask]^{\ast} Y_{\ell m} \, , \\
  \label{eq:mask_w_tilde}
  \widetilde{W} &= \sum_{\ell m} (-1)^{m} \alm[][][\mask]^{\ast}
  Y_{\ell m} \, .
}
A closer analysis of the numerical complexity associated with the
evaluation of \eq{eq:ssff_cut_exact} reveals its significant advantage
over the original \eq{eq:ssff_cut_exact_alm}: we derive the result
exclusively by a series of simple multiplications (either in real
space or in Fourier space), followed by a change of basis via standard
spherical harmonic synthesis or analysis steps, for which fast
numerical libraries are available \citep[e.g.,][]{2005ApJ...622..759G,
  2010ApJS..189..255H, 2011A&A...526A.108R, 2013A&A...554A.112R,
  GGGE:GGGE20071}. Hence, it can be computed in a mathematically exact
way in only \order{\lmax^3} operations.

In practice, we evaluate \eq{eq:ssff_cut_exact} as follows. First,
using all maps in their pixel space representation, we multiply the
cut-sky template \template with an additional instance of the mask,
modified as described by \eq{eq:mask_w_bar}, and transform the result
into spherical harmonic basis. Then, after the coefficients of the
resulting map have been multiplied by the signal power spectrum and a
phase factor, the result is transformed back into real space. Next, we
compute the product of this map with another modified version of the
mask, given by \eq{eq:mask_w_tilde}, and again transform it to Fourier
space. We then obtain the final result by multiplying its spherical
harmonic coefficients with the template and another phase factor.

Defining $\cl[]{X} = \sum_m \frac{1}{2\ell + 1} \langle
\alm[][][\signal] \ (\signal^{\dagger} \template)
\alm[][][\template]^{\ast} \rangle$ as the power spectrum coefficients
computed from \eq{eq:ssff_cut_exact}, for the bias on the cut-sky we
derive
\equ[eq:cutsky_bias]{
b_{\ell} = - \frac{2 \cl[]{X}}{\sum_{\ell^{\prime}} (2 \ell^{\prime} +
  1) \clh[^{\prime}]{\template}} + \frac{\left( \sum_{\ell^{\prime}} (2
  \ell^{\prime} + 1) \cl[^{\prime}]{X} \right) \clh[]{\template}}{\left(
  \sum_{\ell^{\prime}} (2 \ell^{\prime} + 1)
  \clh[^{\prime}]{\template} \right)^2} \, .
}
We obtain the final result by correcting for the limited sky fraction
available to the analysis using the inverse coupling matrix,
\eq{eq:pcl_mask_convolution},
\equ[eq:bias_mask_deconvolve]{
  \overline{b}_{\ell} = \sum_{\ell^{\prime}} M_{\ell \ell^{\prime}}^{-1}
  \, b_{\ell^{\prime}},
}
where $\overline{b}_{\ell}$ is the bias of the mask deconvolved power
spectra.

As already mentioned in \sect{sec:sub_full_sky_single}, the evaluation
of \eq{eq:cutsky_bias} requires knowledge of the unbiased signal power
spectrum, necessitating either the use of a prior on $\cls$ or the
iterative computation of $b_{\ell}$. The convergence of iterative
schemes can be monitored straightforwardly by keeping track of
relative changes in the results of two subsequent iterations. As soon
as this change becomes small compared to, for example, some fraction
of the estimated power spectrum error bar, the algorithm can safely be
terminated.

\subsection{Cut-sky analysis, multiple templates}
\label{sec:sub_cut_sky_multi}

We finally consider the most general case of mode projection with
multiple templates on the cut sky. Building on the results obtained in
the last sections, we start with redefining the normalization matrix
$\left(\templatev^{\dagger} \templatev \right)$. Using
\eq{eq:norm_factor}, we now compute it from template pseudo-power
spectra that are uncorrected for the effect of the mask. Following the
procedure detailed in \sect{sec:sub_full_sky_multi}, we further
introduce the power spectrum
\equ{
  \cl{X^i \times X^j} = \sum_{m} \frac{1}{2 \ell + 1} \langle
  \alm[][][\signal] \ (\signal^{\dagger} \template^i)
  \alm[][][\template]^{j \, \ast} \rangle \, ,
}
an expression that can be straightforwardly computed from
\eq{eq:ssff_cut_exact} using two different templates as inputs. We
then derive a mathematically exact solution for the bias in the most
general case, finding
\eqm[eq:cutskymulti_bias]{
  b_{\ell} = -2 \sum_{i j} \left(\templatev^{\dagger} \templatev
  \right)^{-1}_{i j} \cl{X^j \times X^i} + \sum_{\substack{i j \\ h k}}
  \left(\templatev^{\dagger} \templatev \right)^{-1}_{i j}
  \left(\templatev^{\dagger} \templatev \right)^{-1}_{h k} \\
  \times \left(\sum_{\ell^{\prime}} \llpo[^{\prime}]
  \cl[^{\prime}]{X^j \times X^k} \right) \clh{\template^i \times
    \template^h} \, .
}
The above \eq{eq:cutskymulti_bias} is the main result of this
paper. As discussed in the previous paragraph, this bias estimate must
still be corrected for the limited sky coverage
(\eq{eq:bias_mask_deconvolve}).

\section{Discussion and verification}
\label{sec:verify}

After deriving the analytical expressions to integrate mode projection
into the framework of PCL power spectrum estimation, we now use
simulations to verify our results and assess the scaling of the bias
correction for different input parameters.

\subsection{Signal power spectrum}

Already in the simplistic case where a single template is projected on
the full sky, it is instructive to determine the behaviour of the bias
term for different input power spectra. Drawing a Gaussian realization
of a template from a flat power spectrum, $\cl{\template} =
\mathrm{const.}$, we show results of a power spectrum analysis with
mode projection of 1000 Gaussian signal simulations for two different
cases where $\cls \propto (\ell + 1)^{\{0, -2\}}$ in
\fig{fig:input_cl}. We plot the average relative difference of power
spectra estimated with and without mode projection, an expression
where most of the sample variance cancels. Numerical results agree
well with our analytical bias calculation for both sets of
simulations, demonstrating that it can be reliably removed to obtain
unbiased PCL power spectrum estimates.

As expected, for a flat signal power spectrum we observe a small
negative bias that is constant. Its level can be understood
intuitively: recalling that we have a single degree of freedom (the
template amplitude) that allows the removal of power from one of a
total of $(\lmax + 1)^2$ Fourier modes of the data map, we expect a
bias at a level of $1/(\lmax + 1)^2 \approx 6 \times 10^{-3} \, \%$
for $\lmax = 128$, in agreement with simulations. This picture
changes, however, for a signal that predominantly contains power at a
limited number of multipoles. For a red signal power spectrum, mode
projection mainly removes power on large scales. In this case, we
observe two qualitatively different regimes. While the bias is
negative where the signal is strongest, it turns positive towards
higher multipole moments. Here, we observe a power transfer, where
fluctuations from the template used in the cleaning procedure are
imprinted on the cleaned signal map.

We note in passing that a similar behaviour is expected in simple
component separation methods used for the analysis of CMB data, where
observations at different frequencies are linearly combined to remove
foreground contaminants (e.g., \citealt{1992ApJ...396L...7B,
  2003ApJS..148...97B, 2004ApJ...612..633E}, see also the discussion
in \citealt{2007ApJS..170..288H, 2008PhRvD..78b3003S}).

Owing to the numerical efficiency of the scheme, high-resolution data
sets can be readily analyzed on commodity desktop computers. In
\fig{fig:input_lmax}, we plot the results of 1000 simulations where we
increased the band limit to $\lmax = 2048$, representative for typical
cosmological data sets. In this setting, the full analysis of a single
data set takes less than one wall clock minute on an Intel E5-2687W
processor with eight CPU cores. Projecting a single template on the
full sky for an input power spectrum $\cls \propto (\ell + 1)^{-2}$,
we observe a reduced bias compared to our reference analysis ($\lmax =
128$), following from the larger total number of independent Fourier
modes in the data.

\twofig{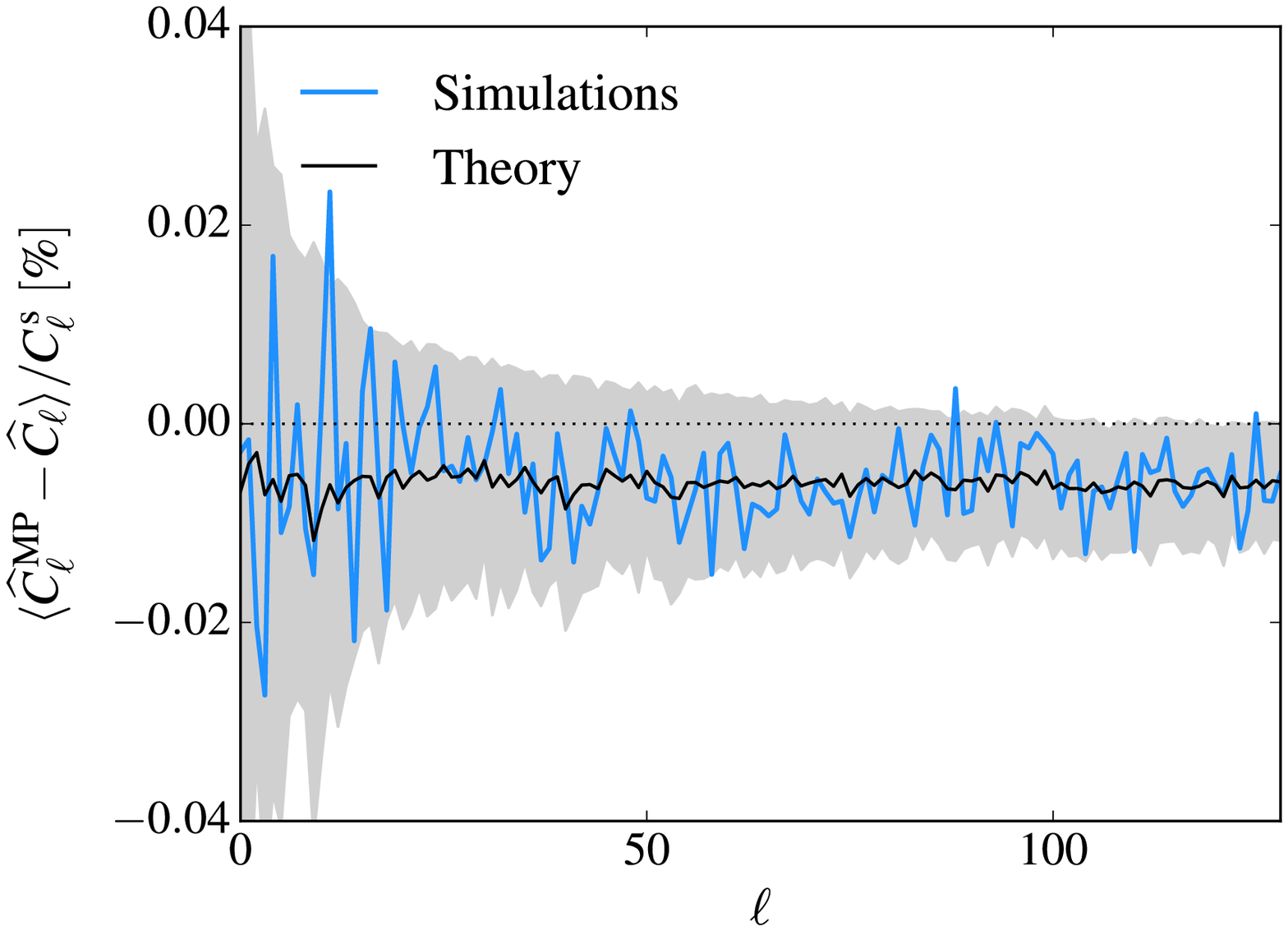}{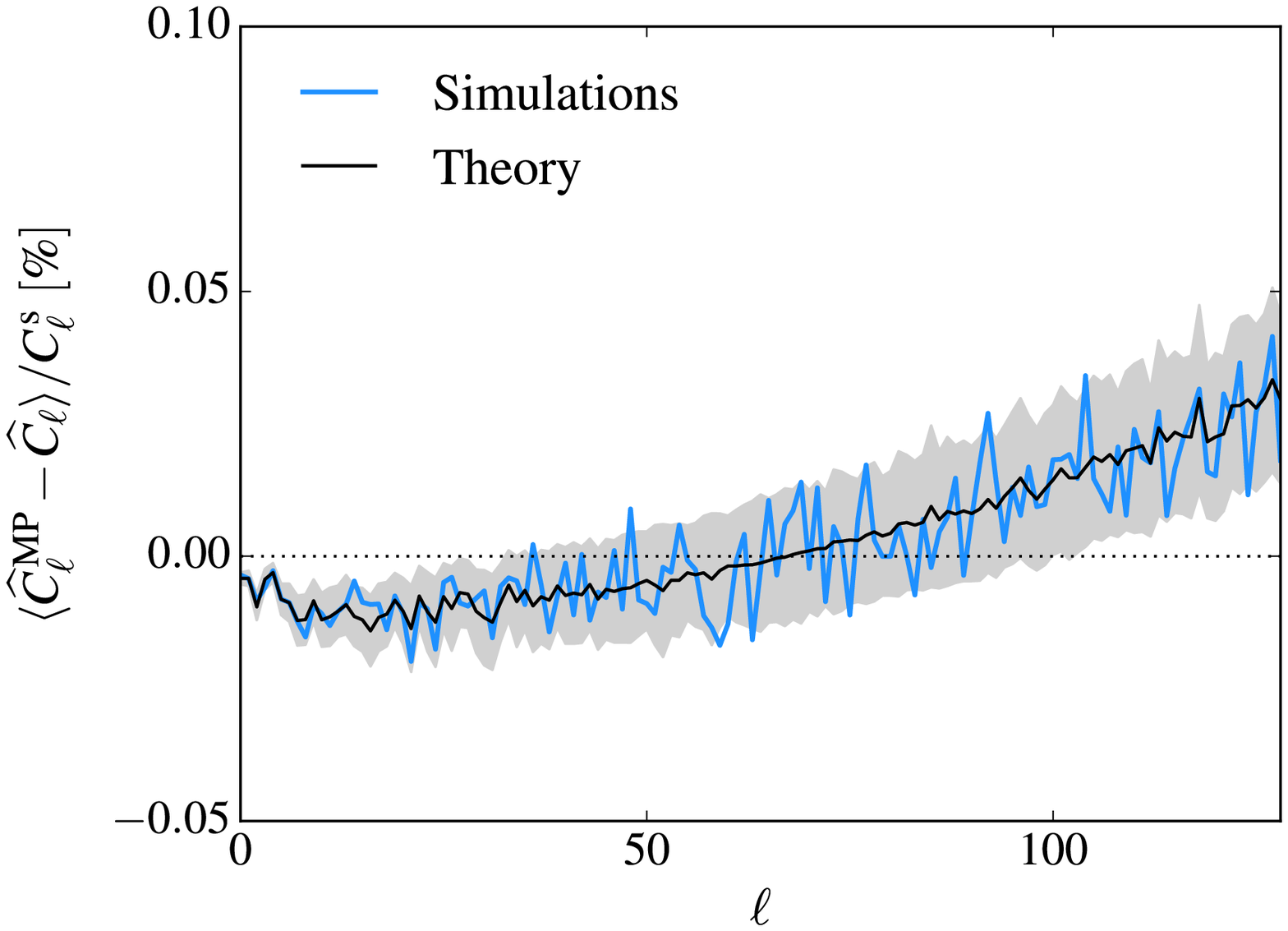}{fig:input_cl}
{The bias introduced by mode projection shows a non-trivial dependence
  on the shape of signal and template power spectra. \emph{Left-hand
    panel:} Result for a flat signal power spectrum, $\cls =
  \mathrm{const.}$ \emph{Right-hand panel:} Bias comparison for a red
  signal power spectrum, $\cls \propto (\ell + 1)^{-2}$. The gray
  regions indicate the empirical $2\text{-}\sigma$ standard error of
  the mean as derived from the simulations.}

\onefig{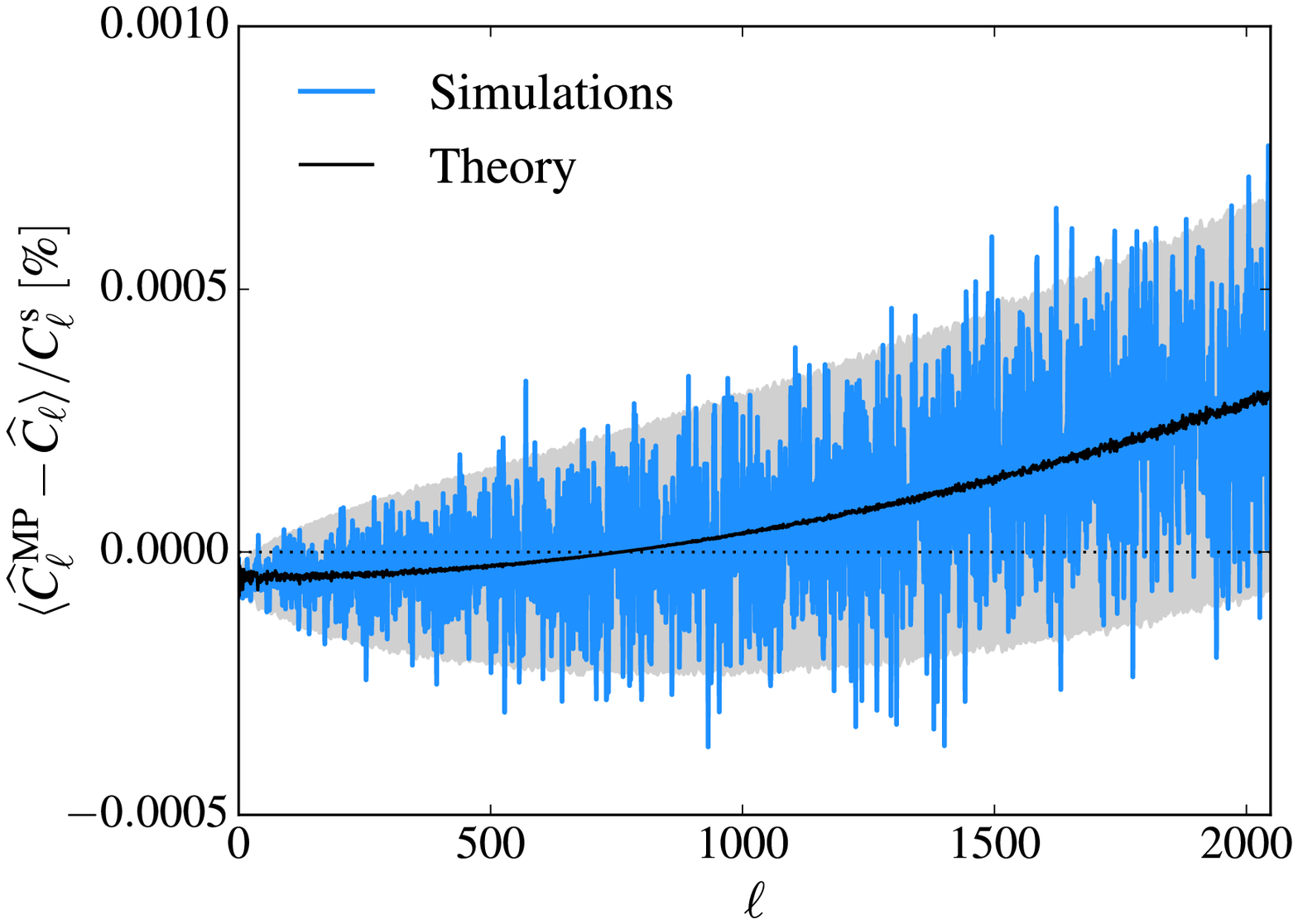}{fig:input_lmax}
{The scheme is efficient enough to be applicable to high-resolution
  data sets. Same as \fig{fig:input_cl}, but plotting results for an
  increased maximum multipole moment of $\lmax = 2048$, derived on an
  off-the-shelf desktop computer.}

\subsection{Number of templates}

Still considering a full sky analysis, we now test the scaling
behaviour of the PCL power spectrum bias induced by mode projection
with the number of templates used in the cleaning procedure. To this
end we repeated the analysis of 1000 simulated realizations of the
data set, drawn from $\cls \propto (\ell + 1)^{-2}$, which we have now
cleaned with 100 randomly generated template maps. The result is shown
in \fig{fig:multi}; compared to the single template case, we observe a
bias that is larger by two orders of magnitude. In case they are not
or only mildly correlated, we indeed expect to see an approximately
linear scaling with the number of templates, since the independent
estimation of the cleaning amplitudes allows the removal of power in
one Fourier mode per template. This observation is of particular
relevance to current and next generation surveys since a robust
analysis may require the projection of the order of hundreds or
thousands of templates. A reliable means to correct for a potentially
large resulting bias is therefore paramount.

\onefig{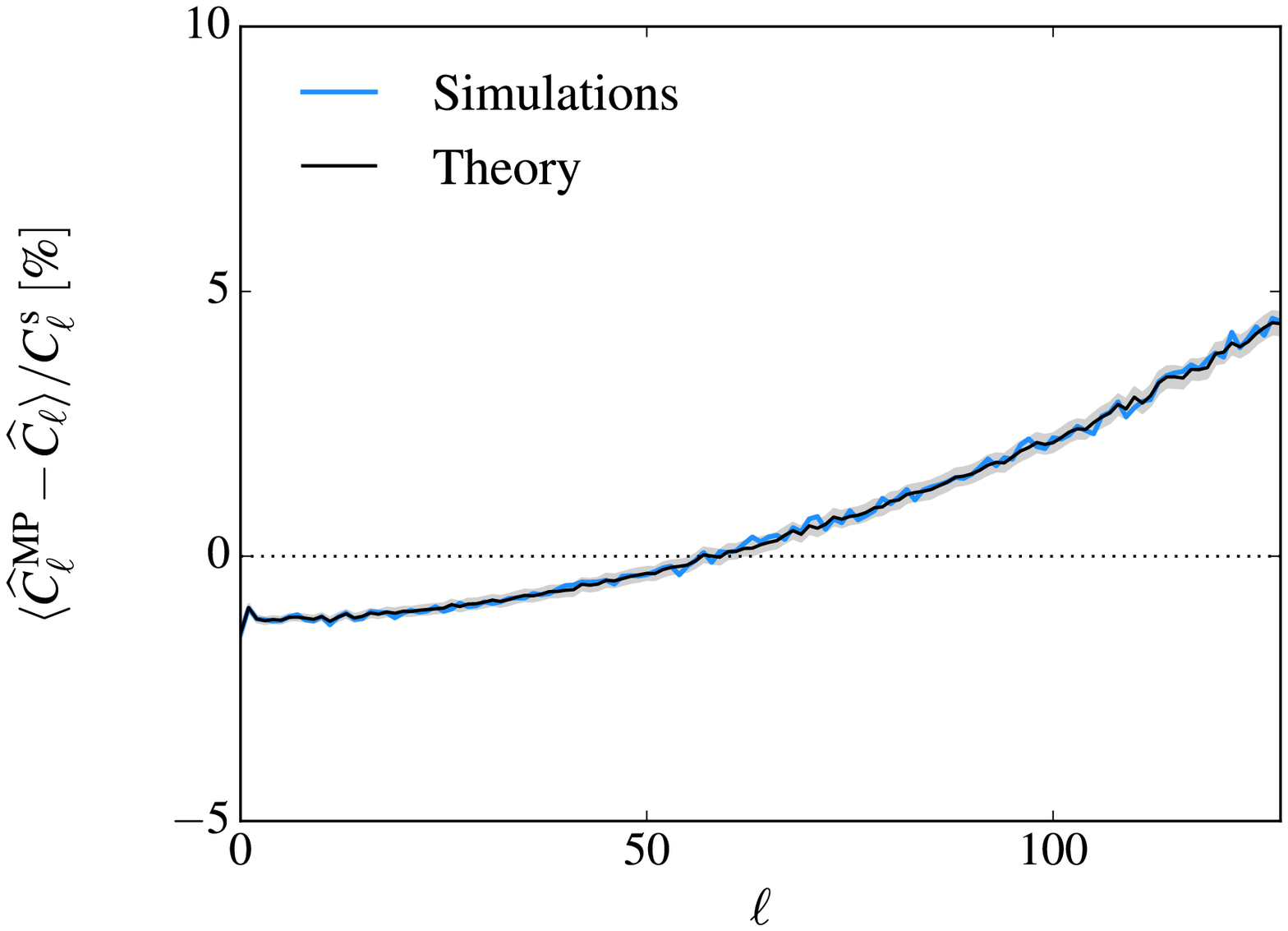}{fig:multi}
{The bias increases approximately linearly with the number of
  templates projected. Same as \fig{fig:input_cl}, but now projecting
  100 instead of a single template. The observed bias becomes larger
  by about two orders of magnitude.}

\subsection{Sky fraction}

In a further set of tests, we probe the impact of a limited sky
fraction used for the analysis on the bias of the power spectra
computed with template projection. In the left-hand panel of
\fig{fig:fsky_multi}, we show the bias for a cut-sky analysis
restricted to $\fsky = 1 \%$. For large to intermediate sky fractions,
we observe a scaling approximately proportional to $1/\fsky$. We note
that for small sky fractions however, this simplified relationship is
expected to break down. In the right-hand panel of
\fig{fig:fsky_multi}, we remove the contribution of 100 templates
while simultaneously restricting the analysis to $\fsky = 1 \%$. In
that case, the relative bias can become larger than unity. The
agreement between simulations and analytical calculation remains good.

\twofig{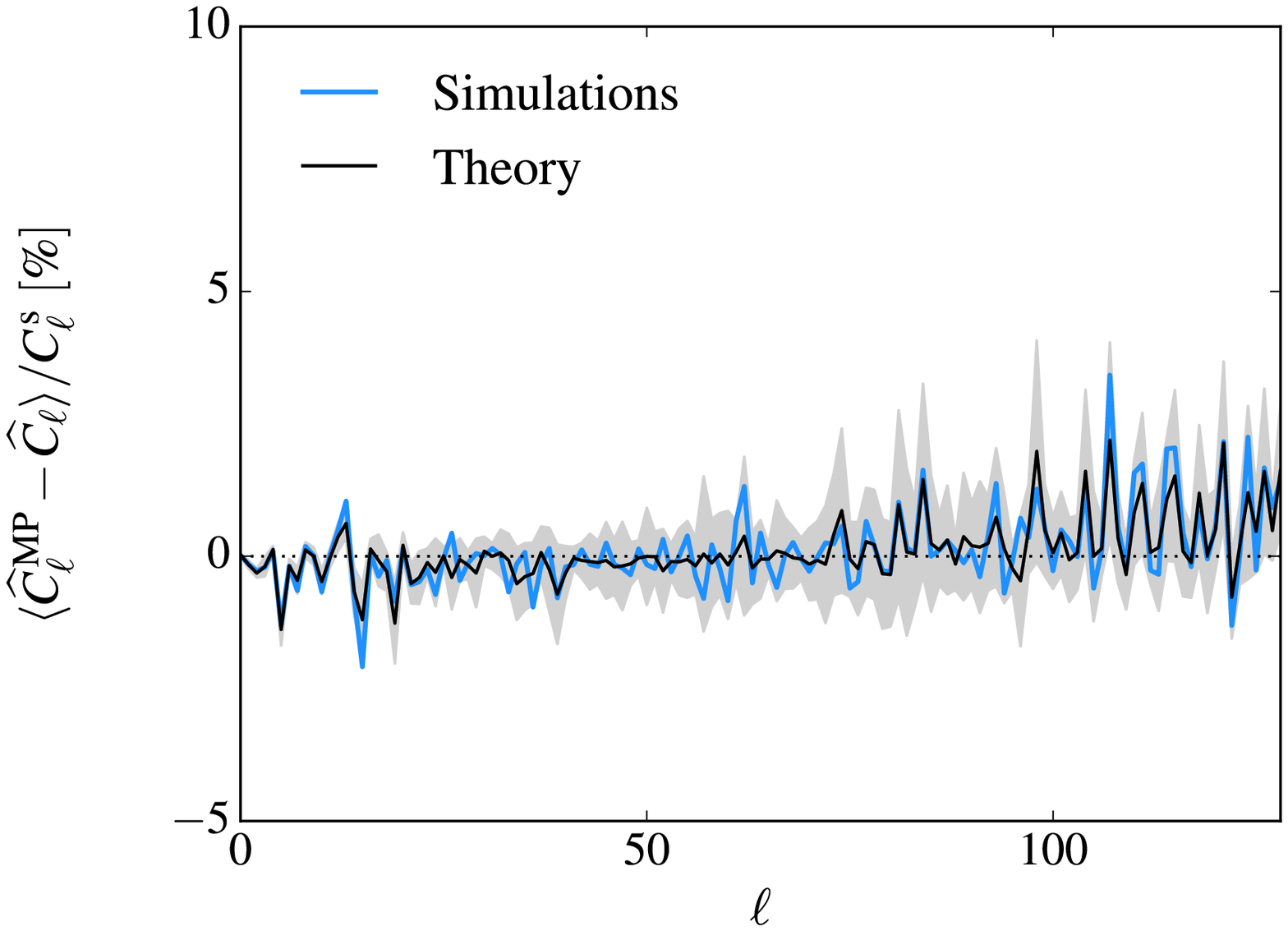}{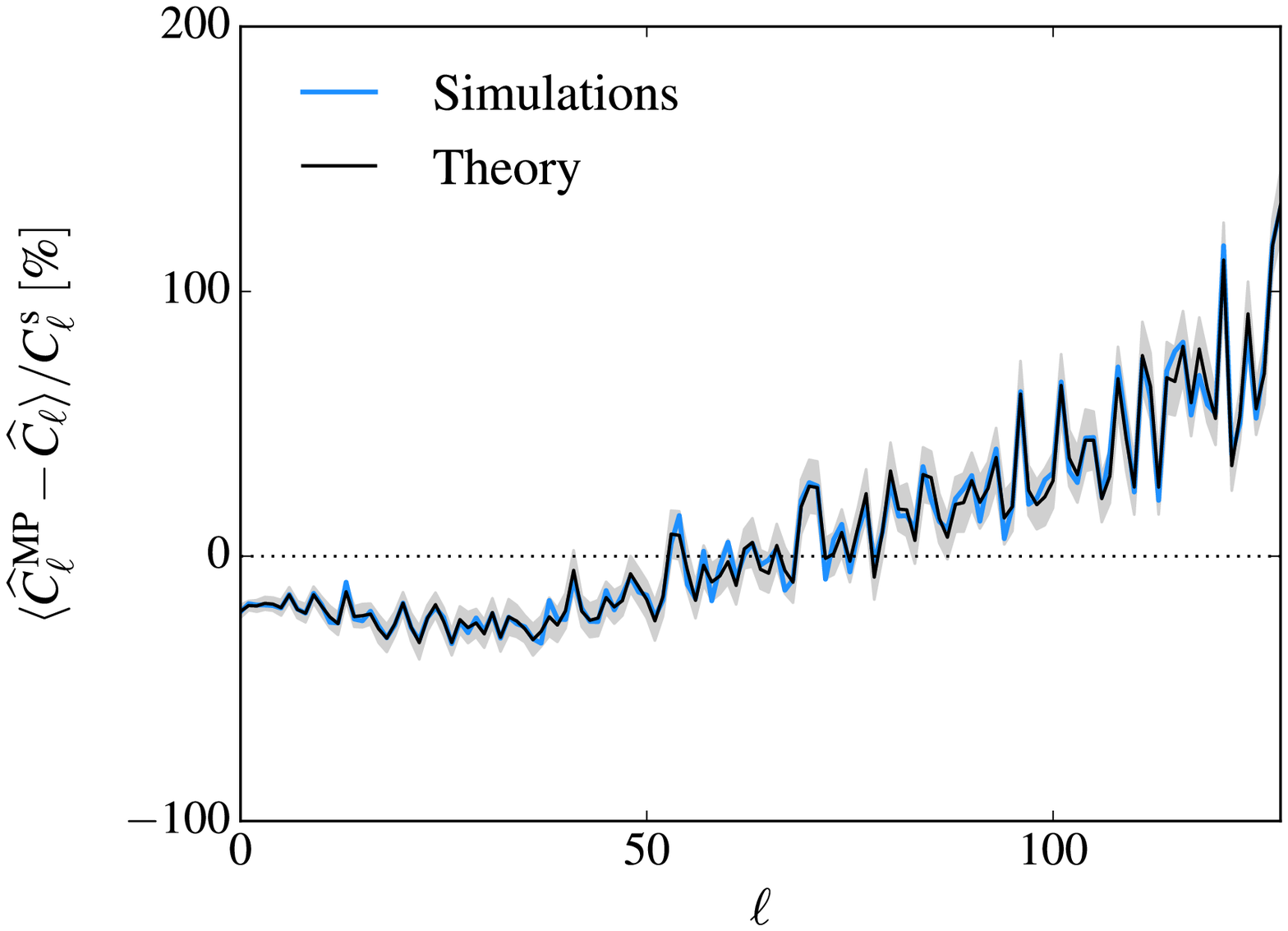}{fig:fsky_multi}
{\emph{Left-hand panel:} The bias is approximately inversely
  proportional to the sky fraction available to the analysis. Same as
  \fig{fig:input_cl}, but now restricting the analysis to $\fsky =
  1\%$. The larger sample variance leads to an increased
  scatter. \emph{Right-hand panel:} Projecting a large number of
  templates on a comparatively small sky area can lead to bias values
  in excess of unity. Same as \fig{fig:input_cl}, but marginalizing
  over 100 templates on $\fsky = 1\%$ of the sky.}

\subsection{Estimator variance}

Implementing mode projection into PCL alters the covariance properties
of power spectrum estimates. While a full analysis is beyond the scope
of this paper, we provide a qualitative assessment of changes in the
estimator variance. In general, the statistical properties of
estimates can be characterized using an analytical description,
simulations, or resampling methods like bootstrapping. Here, we
analyzed the empirical variance of $100\,000$ full-sky power spectra,
computed from signal simulations drawn from $\cls \propto (\ell +
1)^{-2}$. We directly compared the debiased results obtained from maps
that have been cleaned by a single template on the one hand, and the
power spectra computed without mode projection on the other hand. In
general, we find an increased variance with a multipole dependence
resembling the general shape of the bias discussed in the last
paragraphs. Interestingly, as visualized in \fig{fig:variance},
changes in the variance depend on the details of the debiasing
procedure. As mentioned in \sect{sec:theory}, the analytical
expression used to debias the results may depend in a non-trivial way
on the signal power spectrum, leaving us with two options to
proceed. It is possible to either use the current (biased) signal
estimate $\clh{\signal}$ for an iterative correction, or to assume a
prior power spectrum $\cls$ in the calculation. While both approaches
lead to unbiased signal power spectrum estimates in the ensemble
average, the estimator variance will be different. The additional
information introduced by a prior results in a deterministic bias
correction, independent of the signal realization, that in turn leads
to a decreased estimator variance in multipole regions that are most
effectively cleaned.

For flat signal and template power spectra, we can provide an
order-of-magnitude estimate of the expected increase in variance of
iteratively-debiased signal power spectra. Considering the number of
modes removed by projecting $n$ templates, we obtain a rough estimate
of the variance ratio of power spectra computed with and without mode
projection,
\equ{
\left. \mathrm{Var}\left( \clh{\signal, \mathrm{MP}} \right) \middle/
\mathrm{Var}\left( \clh{\signal} \right) \right. \sim \frac{2 (\lmax +
  1)^2}{\fsky \left[ (\lmax + 1)^2 - n \right]} \, ,
}
where $\fsky$ is the sky fraction used in the analysis.

\twofig{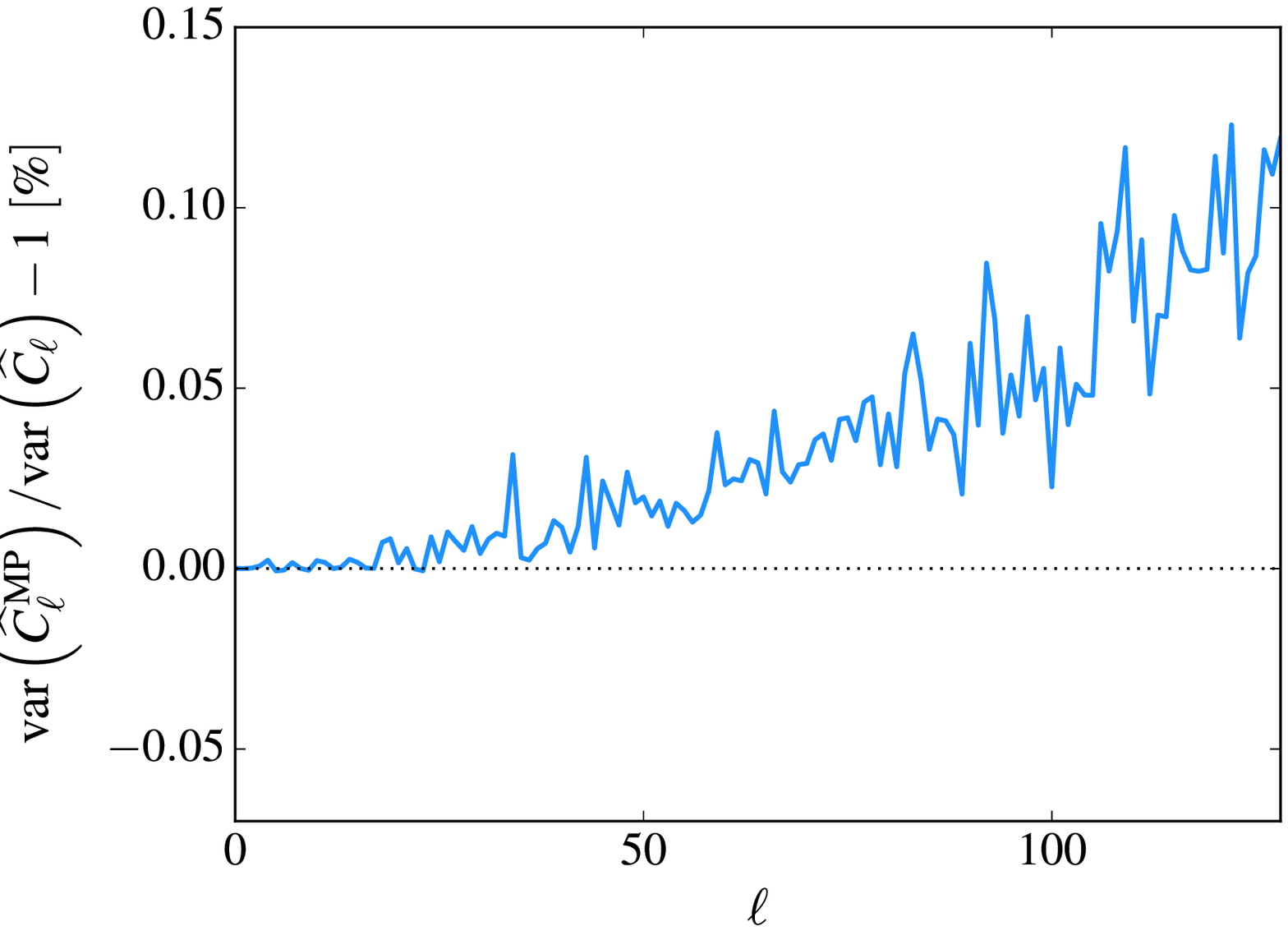}{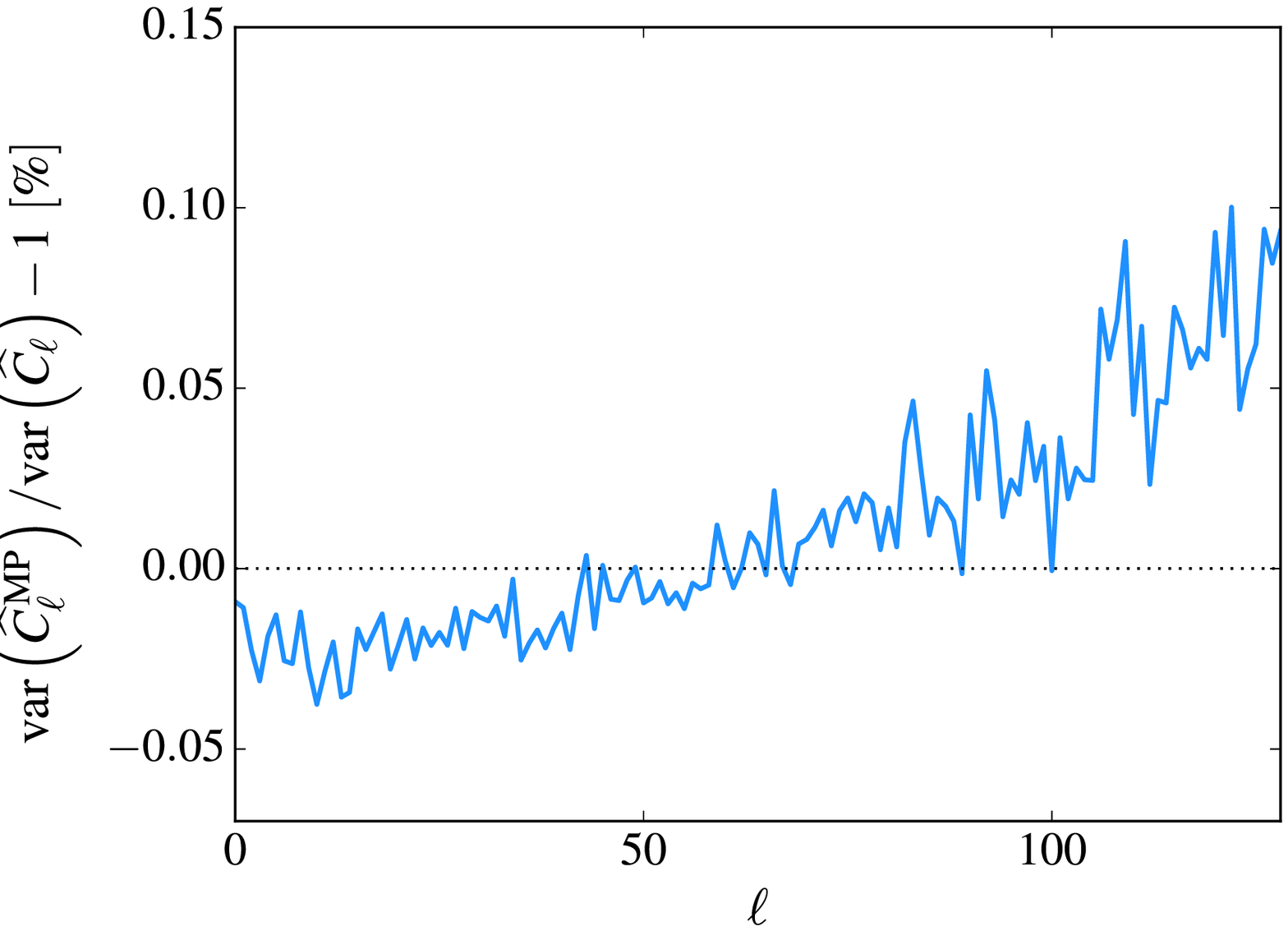}{fig:variance}
{Mode projection induces changes in the estimator variance that depend
  on the details of the bias removal. Debiasing power spectrum
  estimates iteratively increases the estimator variance on all scales
  (\emph{left-hand panel}), while the use of a prior power spectrum
  can result in a multipole range with reduced variance
  (\emph{right-hand panel}).}

\section{Summary and conclusions}
\label{sec:conclusions}

In modern cosmology, two-point correlation function measurements play
a fundamental role in constraining theoretical models with
observational data. In practical application, however, extracting
statistical information about cosmological signals is often hampered
by the presence of contaminants. As a consequence, a number of
strategies have been developed to mitigate their impact on the
scientific analysis. Here, we focus on mode projection, an algorithm
that allows one to marginalize over templates constructed to describe
the spatial patterns of possible systematic effects
\citep{1992ApJ...398..169R}. While it can be straightforwardly
implemented into optimal methods, the application to the popular \pcl
(PCL) estimator, so far, has remained elusive.

In this paper, we have developed a framework to integrate mode
projection into PCL estimation algorithms. We have shown that a naive
projection of templates in general leads to biased power spectrum
estimates. Based on a rigorous mathematical treatment, we then derived
exact closed-form equations for the estimator bias. Recasting the
analytical expressions allowed us to compute them efficiently, thereby
preserving the overall \order{\lmax^3} time complexity of PCL
algorithms.

Applied to a large number of simulations with various input
parameters, we have systematically studied the impact of mode
projection on PCL power spectrum estimates. We identified a nontrivial
dependence of the cleaning procedure on the shape of signal and
template power spectra. We further studied the scaling of the bias
with the band limit of the maps, number of templates projected, and
sky fraction available to the analysis, and discussed the impact of
mode projection on the covariance properties of the power spectrum
estimates. In all cases, we found a good agreement between the bias
observed in simulations and our analytical prediction. We conclude
that the framework presented here allows for a reliable correction of
power spectrum estimates to obtain unbiased results. Possible future
extensions of the algorithm include the generalization to spin-2
fields to allow more robust measurements of, for example, the cosmic
shear signal \citep[e.g.,][]{2000MNRAS.318..625B, 2000astro.ph..3338K,
  2000Natur.405..143W, 2012ApJ...761...15L, 2013MNRAS.430.2200K,
  2015MNRAS.454.3500K, 2016PhRvD..94b2002B}, or the CMB polarization
power spectrum \citep[e.g.,][]{2002Natur.420..772K, 2014PhRvL.112x1101B,
  2014JCAP...10..007N, 2014ApJ...794..171T, 2016A&A...594A..11P}.

Effective strategies for systematics mitigation are instrumental to
fully exploring the information content of ongoing and future
large scale structure surveys like the Sloan Digital Sky Survey
\citep{2000AJ....120.1579Y}, the Dark Energy Survey
\citep{2013AAS...22133501F}, or observations planned with the Dark
Energy Spectroscopic Instrument \citep{2013arXiv1308.0847L}, or the
Large Synoptic Survey Telescope \citep{2009arXiv0912.0201L}. The
results of our studies indicate that the combination of mode
projection and \pcl power spectrum estimation offers an attractive
means to robustly measure the two-point correlation function in the
presence of contaminants, an important milestone on the way to
reliable clustering estimates.

\section*{Acknowledgements}

We are grateful to our referee, An\v{z}e Slosar, for useful
discussions and thank Andrew Pontzen for commenting on the draft. FE,
BL, and HVP were partially supported by the European Research Council
under the European Community's Seventh Framework Programme
(FP7/2007-2013) / ERC grant agreement no 306478-CosmicDawn. Some of
the results in this paper have been derived using the
HEALPix\footnote{\url{http://healpix.sourceforge.net}}
\citep{2005ApJ...622..759G} package.

\bibliographystyle{mnras}
\bibliography{literature}

\begin{thebibliography}{}
\makeatletter
\relax
\def\mn@urlcharsother{\let\do\@makeother \do\$\do\&\do\#\do\^\do\_\do\%\do\~}
\def\mn@doi{\begingroup\mn@urlcharsother \@ifnextchar [ {\mn@doi@}
  {\mn@doi@[]}}
\def\mn@doi@[#1]#2{\def\@tempa{#1}\ifx\@tempa\@empty \href
  {http://dx.doi.org/#2} {doi:#2}\else \href {http://dx.doi.org/#2} {#1}\fi
  \endgroup}
\def\mn@eprint#1#2{\mn@eprint@#1:#2::\@nil}
\def\mn@eprint@arXiv#1{\href {http://arxiv.org/abs/#1} {{\tt arXiv:#1}}}
\def\mn@eprint@dblp#1{\href {http://dblp.uni-trier.de/rec/bibtex/#1.xml}
  {dblp:#1}}
\def\mn@eprint@#1:#2:#3:#4\@nil{\def\@tempa {#1}\def\@tempb {#2}\def\@tempc
  {#3}\ifx \@tempc \@empty \let \@tempc \@tempb \let \@tempb \@tempa \fi \ifx
  \@tempb \@empty \def\@tempb {arXiv}\fi \@ifundefined
  {mn@eprint@\@tempb}{\@tempb:\@tempc}{\expandafter \expandafter \csname
  mn@eprint@\@tempb\endcsname \expandafter{\@tempc}}}

\bibitem[\protect\citeauthoryear{{Awan} et~al.,}{{Awan}
  et~al.}{2016}]{2016ApJ...829...50A}
{Awan} H.,  et~al., 2016, \mn@doi [\apj] {10.3847/0004-637X/829/1/50}, 829, 50

\bibitem[\protect\citeauthoryear{{BICEP2 Collaboration} et~al.,}{{BICEP2
  Collaboration} et~al.}{2014}]{2014PhRvL.112x1101B}
{BICEP2 Collaboration} et~al., 2014, \mn@doi [Physical Review Letters]
  {10.1103/PhysRevLett.112.241101}, 112, 241101

\bibitem[\protect\citeauthoryear{{BICEP2/Keck and Planck Collaborations}
  et~al.,}{{BICEP2/Keck and Planck Collaborations}
  et~al.}{2015}]{2015PhRvL.114j1301B}
{BICEP2/Keck and Planck Collaborations} et~al., 2015, \mn@doi [Physical Review
  Letters] {10.1103/PhysRevLett.114.101301}, 114, 101301

\bibitem[\protect\citeauthoryear{{Bacon}, {Refregier}  \& {Ellis}}{{Bacon}
  et~al.}{2000}]{2000MNRAS.318..625B}
{Bacon} D.~J.,  {Refregier} A.~R.,   {Ellis} R.~S.,  2000, \mn@doi [\mnras]
  {10.1046/j.1365-8711.2000.03851.x}, 318, 625

\bibitem[\protect\citeauthoryear{{Becker} et~al.,}{{Becker}
  et~al.}{2016}]{2016PhRvD..94b2002B}
{Becker} M.~R.,  et~al., 2016, \mn@doi [\prd] {10.1103/PhysRevD.94.022002}, 94,
  022002

\bibitem[\protect\citeauthoryear{{Bennett} et~al.,}{{Bennett}
  et~al.}{1992}]{1992ApJ...396L...7B}
{Bennett} C.~L.,  et~al., 1992, \mn@doi [\apjl] {10.1086/186505}, 396, L7

\bibitem[\protect\citeauthoryear{{Bennett} et~al.,}{{Bennett}
  et~al.}{2003}]{2003ApJS..148...97B}
{Bennett} C.~L.,  et~al., 2003, \mn@doi [\apjs] {10.1086/377252}, 148, 97

\bibitem[\protect\citeauthoryear{{Beutler} et~al.,}{{Beutler}
  et~al.}{2011}]{2011MNRAS.416.3017B}
{Beutler} F.,  et~al., 2011, \mn@doi [\mnras]
  {10.1111/j.1365-2966.2011.19250.x}, 416, 3017

\bibitem[\protect\citeauthoryear{{Blake} \& {Wall}}{{Blake} \&
  {Wall}}{2002}]{2002MNRAS.329L..37B}
{Blake} C.,  {Wall} J.,  2002, \mn@doi [\mnras]
  {10.1046/j.1365-8711.2002.05163.x}, 329, L37

\bibitem[\protect\citeauthoryear{{Bond}, {Jaffe}  \& {Knox}}{{Bond}
  et~al.}{1998}]{1998PhRvD..57.2117B}
{Bond} J.~R.,  {Jaffe} A.~H.,   {Knox} L.,  1998, \prd, 57, 2117

\bibitem[\protect\citeauthoryear{{Borrill}}{{Borrill}}{1999}]{1999PhRvD..59b7302B}
{Borrill} J.,  1999, \mn@doi [\prd] {10.1103/PhysRevD.59.027302}, 59, 027302

\bibitem[\protect\citeauthoryear{{Coil} et~al.,}{{Coil}
  et~al.}{2008}]{2008ApJ...672..153C}
{Coil} A.~L.,  et~al., 2008, \mn@doi [\apj] {10.1086/523639}, 672, 153

\bibitem[\protect\citeauthoryear{{Crocce} et~al.,}{{Crocce}
  et~al.}{2016}]{2016MNRAS.455.4301C}
{Crocce} M.,  et~al., 2016, \mn@doi [\mnras] {10.1093/mnras/stv2590}, 455, 4301

\bibitem[\protect\citeauthoryear{{Croom} et~al.,}{{Croom}
  et~al.}{2005}]{2005MNRAS.356..415C}
{Croom} S.~M.,  et~al., 2005, \mn@doi [\mnras]
  {10.1111/j.1365-2966.2004.08379.x}, 356, 415

\bibitem[\protect\citeauthoryear{{Edmonds}}{{Edmonds}}{1996}]{edmonds1996angular}
{Edmonds} A.~R.,  1996, Angular Momentum in Quantum Mechanics.
Investigations in Physics Series, Princeton University Press, \url
  {http://press.princeton.edu/titles/478.html}

\bibitem[\protect\citeauthoryear{{Eisenstein} et~al.,}{{Eisenstein}
  et~al.}{2005}]{2005ApJ...633..560E}
{Eisenstein} D.~J.,  et~al., 2005, \mn@doi [\apj] {10.1086/466512}, 633, 560

\bibitem[\protect\citeauthoryear{{Elsner} \& {Wandelt}}{{Elsner} \&
  {Wandelt}}{2013}]{2013A&A...549A.111E}
{Elsner} F.,  {Wandelt} B.~D.,  2013, \mn@doi [\aap]
  {10.1051/0004-6361/201220586}, 549, A111

\bibitem[\protect\citeauthoryear{{Elsner}, {Leistedt}  \& {Peiris}}{{Elsner}
  et~al.}{2016}]{2016MNRAS.456.2095E}
{Elsner} F.,  {Leistedt} B.,   {Peiris} H.~V.,  2016, \mn@doi [\mnras]
  {10.1093/mnras/stv2777}, 456, 2095

\bibitem[\protect\citeauthoryear{{Eriksen}, {Banday}, {G{\'o}rski}  \&
  {Lilje}}{{Eriksen} et~al.}{2004}]{2004ApJ...612..633E}
{Eriksen} H.~K.,  {Banday} A.~J.,  {G{\'o}rski} K.~M.,   {Lilje} P.~B.,  2004,
  \mn@doi [\apj] {10.1086/422807}, 612, 633

\bibitem[\protect\citeauthoryear{{Feldman}, {Kaiser}  \& {Peacock}}{{Feldman}
  et~al.}{1994}]{1994ApJ...426...23F}
{Feldman} H.~A.,  {Kaiser} N.,   {Peacock} J.~A.,  1994, \mn@doi [\apj]
  {10.1086/174036}, 426, 23

\bibitem[\protect\citeauthoryear{{Fowler} et~al.,}{{Fowler}
  et~al.}{2010}]{2010ApJ...722.1148F}
{Fowler} J.~W.,  et~al., 2010, \mn@doi [\apj] {10.1088/0004-637X/722/2/1148},
  722, 1148

\bibitem[\protect\citeauthoryear{{Frieman} \& {Dark Energy Survey
  Collaboration}}{{Frieman} \& {Dark Energy Survey
  Collaboration}}{2013}]{2013AAS...22133501F}
{Frieman} J.,  {Dark Energy Survey Collaboration} 2013, in American
  Astronomical Society Meeting Abstracts \#221. p. 335.01, \url
  {http://adsabs.harvard.edu/abs/2013AAS...22133501F}

\bibitem[\protect\citeauthoryear{{Gaunt}}{{Gaunt}}{1929}]{Gaunt151}
{Gaunt} J.~A.,  1929, \mn@doi [Royal Society of London Philosophical
  Transactions Series A] {10.1098/rsta.1929.0004}, 228, 151

\bibitem[\protect\citeauthoryear{{G{\'o}rski}, {Hivon}, {Banday}, {Wandelt},
  {Hansen}, {Reinecke}  \& {Bartelmann}}{{G{\'o}rski}
  et~al.}{2005}]{2005ApJ...622..759G}
{G{\'o}rski} K.~M.,  {Hivon} E.,  {Banday} A.~J.,  {Wandelt} B.~D.,  {Hansen}
  F.~K.,  {Reinecke} M.,   {Bartelmann} M.,  2005, \mn@doi [\apj]
  {10.1086/427976}, 622, 759

\bibitem[\protect\citeauthoryear{{Gundersen} et~al.,}{{Gundersen}
  et~al.}{1995}]{1995ApJ...443L..57G}
{Gundersen} J.~O.,  et~al., 1995, \mn@doi [\apjl] {10.1086/187835}, 443, L57

\bibitem[\protect\citeauthoryear{{Halverson} et~al.,}{{Halverson}
  et~al.}{2002}]{2002ApJ...568...38H}
{Halverson} N.~W.,  et~al., 2002, \mn@doi [\apj] {10.1086/338879}, 568, 38

\bibitem[\protect\citeauthoryear{{Hanany} et~al.,}{{Hanany}
  et~al.}{2000}]{2000ApJ...545L...5H}
{Hanany} S.,  et~al., 2000, \mn@doi [\apjl] {10.1086/317322}, 545, L5

\bibitem[\protect\citeauthoryear{{Hancock}, {Davies}, {Lasenby}, {de La Cruz},
  {Watson}, {Rebolo}  \& {Beckman}}{{Hancock}
  et~al.}{1994}]{1994Natur.367..333H}
{Hancock} S.,  {Davies} R.~D.,  {Lasenby} A.~N.,  {de La Cruz} C.~M.~G.,
  {Watson} R.~A.,  {Rebolo} R.,   {Beckman} J.~E.,  1994, \mn@doi [\nat]
  {10.1038/367333a0}, 367, 333

\bibitem[\protect\citeauthoryear{{Hermit}, {Santiago}, {Lahav}, {Strauss},
  {Davis}, {Dressler}  \& {Huchra}}{{Hermit}
  et~al.}{1996}]{1996MNRAS.283..709H}
{Hermit} S.,  {Santiago} B.~X.,  {Lahav} O.,  {Strauss} M.~A.,  {Davis} M.,
  {Dressler} A.,   {Huchra} J.~P.,  1996, \mn@doi [\mnras]
  {10.1093/mnras/283.2.709}, 283, 709

\bibitem[\protect\citeauthoryear{{Hinshaw} et~al.,}{{Hinshaw}
  et~al.}{2003}]{2003ApJS..148..135H}
{Hinshaw} G.,  et~al., 2003, \mn@doi [\apjs] {10.1086/377225}, 148, 135

\bibitem[\protect\citeauthoryear{{Hinshaw} et~al.,}{{Hinshaw}
  et~al.}{2007}]{2007ApJS..170..288H}
{Hinshaw} G.,  et~al., 2007, \mn@doi [\apjs] {10.1086/513698}, 170, 288

\bibitem[\protect\citeauthoryear{{Hivon}, {G{\'o}rski}, {Netterfield}, {Crill},
  {Prunet}  \& {Hansen}}{{Hivon} et~al.}{2002}]{2002ApJ...567....2H}
{Hivon} E.,  {G{\'o}rski} K.~M.,  {Netterfield} C.~B.,  {Crill} B.~P.,
  {Prunet} S.,   {Hansen} F.,  2002, \mn@doi [\apj] {10.1086/338126}, 567, 2

\bibitem[\protect\citeauthoryear{{Ho} et~al.,}{{Ho}
  et~al.}{2012}]{2012ApJ...761...14H}
{Ho} S.,  et~al., 2012, \mn@doi [\apj] {10.1088/0004-637X/761/1/14}, 761, 14

\bibitem[\protect\citeauthoryear{{Huffenberger} \& {Wandelt}}{{Huffenberger} \&
  {Wandelt}}{2010}]{2010ApJS..189..255H}
{Huffenberger} K.~M.,  {Wandelt} B.~D.,  2010, \mn@doi [\apjs]
  {10.1088/0067-0049/189/2/255}, 189, 255

\bibitem[\protect\citeauthoryear{{Huterer}, {Cunha}  \& {Fang}}{{Huterer}
  et~al.}{2013}]{2013MNRAS.432.2945H}
{Huterer} D.,  {Cunha} C.~E.,   {Fang} W.,  2013, \mn@doi [\mnras]
  {10.1093/mnras/stt653}, 432, 2945

\bibitem[\protect\citeauthoryear{{Kaiser}, {Wilson}  \& {Luppino}}{{Kaiser}
  et~al.}{2000}]{2000astro.ph..3338K}
{Kaiser} N.,  {Wilson} G.,   {Luppino} G.~A.,  2000, ArXiv e-prints,
  astro-ph/0003338

\bibitem[\protect\citeauthoryear{{Kalus}, {Percival}, {Bacon}  \&
  {Samushia}}{{Kalus} et~al.}{2016}]{2016MNRAS.463..467K}
{Kalus} B.,  {Percival} W.~J.,  {Bacon} D.~J.,   {Samushia} L.,  2016, \mn@doi
  [\mnras] {10.1093/mnras/stw2008}, 463, 467

\bibitem[\protect\citeauthoryear{{Kilbinger} et~al.,}{{Kilbinger}
  et~al.}{2013}]{2013MNRAS.430.2200K}
{Kilbinger} M.,  et~al., 2013, \mn@doi [\mnras] {10.1093/mnras/stt041}, 430,
  2200

\bibitem[\protect\citeauthoryear{{Kim} et~al.,}{{Kim}
  et~al.}{2014}]{2014MNRAS.438..825K}
{Kim} J.-W.,  et~al., 2014, \mn@doi [\mnras] {10.1093/mnras/stt2245}, 438, 825

\bibitem[\protect\citeauthoryear{{Kovac}, {Leitch}, {Pryke}, {Carlstrom},
  {Halverson}  \& {Holzapfel}}{{Kovac} et~al.}{2002}]{2002Natur.420..772K}
{Kovac} J.~M.,  {Leitch} E.~M.,  {Pryke} C.,  {Carlstrom} J.~E.,  {Halverson}
  N.~W.,   {Holzapfel} W.~L.,  2002, \mn@doi [\nat] {10.1038/nature01269}, 420,
  772

\bibitem[\protect\citeauthoryear{{Kuijken} et~al.,}{{Kuijken}
  et~al.}{2015}]{2015MNRAS.454.3500K}
{Kuijken} K.,  et~al., 2015, \mn@doi [\mnras] {10.1093/mnras/stv2140}, 454,
  3500

\bibitem[\protect\citeauthoryear{{LSST Science Collaboration} et~al.,}{{LSST
  Science Collaboration} et~al.}{2009}]{2009arXiv0912.0201L}
{LSST Science Collaboration} et~al., 2009, ArXiv e-prints, 0912.0201

\bibitem[\protect\citeauthoryear{{Leistedt} \& {Peiris}}{{Leistedt} \&
  {Peiris}}{2014}]{2014MNRAS.444....2L}
{Leistedt} B.,  {Peiris} H.~V.,  2014, \mn@doi [\mnras]
  {10.1093/mnras/stu1439}, 444, 2

\bibitem[\protect\citeauthoryear{{Leistedt}, {Peiris}, {Mortlock},
  {Benoit-L{\'e}vy}  \& {Pontzen}}{{Leistedt}
  et~al.}{2013}]{2013MNRAS.435.1857L}
{Leistedt} B.,  {Peiris} H.~V.,  {Mortlock} D.~J.,  {Benoit-L{\'e}vy} A.,
  {Pontzen} A.,  2013, \mn@doi [\mnras] {10.1093/mnras/stt1359}, 435, 1857

\bibitem[\protect\citeauthoryear{{Leistedt} et~al.,}{{Leistedt}
  et~al.}{2016}]{2016ApJS..226...24L}
{Leistedt} B.,  et~al., 2016, \mn@doi [\apjs] {10.3847/0067-0049/226/2/24},
  226, 24

\bibitem[\protect\citeauthoryear{{Levi} et~al.,}{{Levi}
  et~al.}{2013}]{2013arXiv1308.0847L}
{Levi} M.,  et~al., 2013, preprint (\mn@eprint {arXiv} {1308.0847})

\bibitem[\protect\citeauthoryear{{Lin} et~al.,}{{Lin}
  et~al.}{2012}]{2012ApJ...761...15L}
{Lin} H.,  et~al., 2012, \mn@doi [\apj] {10.1088/0004-637X/761/1/15}, 761, 15

\bibitem[\protect\citeauthoryear{{Lueker} et~al.,}{{Lueker}
  et~al.}{2010}]{2010ApJ...719.1045L}
{Lueker} M.,  et~al., 2010, \mn@doi [\apj] {10.1088/0004-637X/719/2/1045}, 719,
  1045

\bibitem[\protect\citeauthoryear{{Maller}, {McIntosh}, {Katz}  \&
  {Weinberg}}{{Maller} et~al.}{2005}]{2005ApJ...619..147M}
{Maller} A.~H.,  {McIntosh} D.~H.,  {Katz} N.,   {Weinberg} M.~D.,  2005,
  \mn@doi [\apj] {10.1086/426181}, 619, 147

\bibitem[\protect\citeauthoryear{{Naess} et~al.,}{{Naess}
  et~al.}{2014}]{2014JCAP...10..007N}
{Naess} S.,  et~al., 2014, \mn@doi [\jcap] {10.1088/1475-7516/2014/10/007}, 10,
  007

\bibitem[\protect\citeauthoryear{{Netterfield}, {Devlin}, {Jarosik}, {Page}  \&
  {Wollack}}{{Netterfield} et~al.}{1997}]{1997ApJ...474...47N}
{Netterfield} C.~B.,  {Devlin} M.~J.,  {Jarosik} N.,  {Page} L.,   {Wollack}
  E.~J.,  1997, \apj, 474, 47

\bibitem[\protect\citeauthoryear{{Norberg} et~al.,}{{Norberg}
  et~al.}{2001}]{2001MNRAS.328...64N}
{Norberg} P.,  et~al., 2001, \mn@doi [\mnras]
  {10.1046/j.1365-8711.2001.04839.x}, 328, 64

\bibitem[\protect\citeauthoryear{{Planck Collaboration} et~al.,}{{Planck
  Collaboration} et~al.}{2014}]{2014A&A...571A..15P}
{Planck Collaboration} et~al., 2014, \mn@doi [\aap]
  {10.1051/0004-6361/201321573}, 571, A15

\bibitem[\protect\citeauthoryear{{Planck Collaboration} et~al.,}{{Planck
  Collaboration} et~al.}{2016}]{2016A&A...594A..11P}
{Planck Collaboration} et~al., 2016, \mn@doi [\aap]
  {10.1051/0004-6361/201526926}, \href
  {http://adsabs.harvard.edu/abs/2016A%26A...594A..11P} {594, A11}

\bibitem[\protect\citeauthoryear{{Reid} et~al.,}{{Reid}
  et~al.}{2010}]{2010MNRAS.404...60R}
{Reid} B.~A.,  et~al., 2010, \mn@doi [\mnras]
  {10.1111/j.1365-2966.2010.16276.x}, 404, 60

\bibitem[\protect\citeauthoryear{{Reinecke}}{{Reinecke}}{2011}]{2011A&A...526A.108R}
{Reinecke} M.,  2011, \mn@doi [\aap] {10.1051/0004-6361/201015906}, 526, A108+

\bibitem[\protect\citeauthoryear{{Reinecke} \& {Seljebotn}}{{Reinecke} \&
  {Seljebotn}}{2013}]{2013A&A...554A.112R}
{Reinecke} M.,  {Seljebotn} D.~S.,  2013, \mn@doi [\aap]
  {10.1051/0004-6361/201321494}, 554, A112

\bibitem[\protect\citeauthoryear{{Ross} et~al.,}{{Ross}
  et~al.}{2011}]{2011MNRAS.417.1350R}
{Ross} A.~J.,  et~al., 2011, \mn@doi [\mnras]
  {10.1111/j.1365-2966.2011.19351.x}, 417, 1350

\bibitem[\protect\citeauthoryear{{Ross} et~al.,}{{Ross}
  et~al.}{2012}]{2012MNRAS.424..564R}
{Ross} A.~J.,  et~al., 2012, \mn@doi [\mnras]
  {10.1111/j.1365-2966.2012.21235.x}, 424, 564

\bibitem[\protect\citeauthoryear{{Ross} et~al.,}{{Ross}
  et~al.}{2016}]{2016arXiv160703145R}
{Ross} A.~J.,  et~al., 2016, preprint (\mn@eprint {arXiv} {1607.03145})

\bibitem[\protect\citeauthoryear{{Rybicki} \& {Press}}{{Rybicki} \&
  {Press}}{1992}]{1992ApJ...398..169R}
{Rybicki} G.~B.,  {Press} W.~H.,  1992, \mn@doi [\apj] {10.1086/171845}, 398,
  169

\bibitem[\protect\citeauthoryear{{Saha}, {Prunet}, {Jain}  \&
  {Souradeep}}{{Saha} et~al.}{2008}]{2008PhRvD..78b3003S}
{Saha} R.,  {Prunet} S.,  {Jain} P.,   {Souradeep} T.,  2008, \mn@doi [\prd]
  {10.1103/PhysRevD.78.023003}, 78, 023003

\bibitem[\protect\citeauthoryear{{Schaeffer}}{{Schaeffer}}{2013}]{GGGE:GGGE20071}
{Schaeffer} N.,  2013, \mn@doi [Geochemistry, Geophysics, Geosystems]
  {10.1002/ggge.20071}, 14, 751

\bibitem[\protect\citeauthoryear{{Scranton} et~al.,}{{Scranton}
  et~al.}{2002}]{2002ApJ...579...48S}
{Scranton} R.,  et~al., 2002, \mn@doi [\apj] {10.1086/342786}, 579, 48

\bibitem[\protect\citeauthoryear{{Slosar}, {Seljak}  \& {Makarov}}{{Slosar}
  et~al.}{2004}]{2004PhRvD..69l3003S}
{Slosar} A.,  {Seljak} U.,   {Makarov} A.,  2004, \mn@doi [\prd]
  {10.1103/PhysRevD.69.123003}, 69, 123003

\bibitem[\protect\citeauthoryear{{Smith}, {Senatore}  \& {Zaldarriaga}}{{Smith}
  et~al.}{2009}]{2009JCAP...09..006S}
{Smith} K.~M.,  {Senatore} L.,   {Zaldarriaga} M.,  2009, \mn@doi [\jcap]
  {10.1088/1475-7516/2009/09/006}, 9, 6

\bibitem[\protect\citeauthoryear{{Smoot} et~al.,}{{Smoot}
  et~al.}{1992}]{1992ApJ...396L...1S}
{Smoot} G.~F.,  et~al., 1992, \mn@doi [\apjl] {10.1086/186504}, 396, L1

\bibitem[\protect\citeauthoryear{{Tegmark}}{{Tegmark}}{1997}]{1997PhRvD..55.5895T}
{Tegmark} M.,  1997, \mn@doi [\prd] {10.1103/PhysRevD.55.5895}, 55, 5895

\bibitem[\protect\citeauthoryear{{Tegmark}, {Hamilton}, {Strauss}, {Vogeley}
  \& {Szalay}}{{Tegmark} et~al.}{1998}]{1998ApJ...499..555T}
{Tegmark} M.,  {Hamilton} A.~J.~S.,  {Strauss} M.~A.,  {Vogeley} M.~S.,
  {Szalay} A.~S.,  1998, \apj, 499, 555

\bibitem[\protect\citeauthoryear{{Tegmark} et~al.,}{{Tegmark}
  et~al.}{2004}]{2004ApJ...606..702T}
{Tegmark} M.,  et~al., 2004, \mn@doi [\apj] {10.1086/382125}, 606, 702

\bibitem[\protect\citeauthoryear{{The Polarbear Collaboration: P.~A.~R.~Ade}
  et~al.,}{{The Polarbear Collaboration: P.~A.~R.~Ade}
  et~al.}{2014}]{2014ApJ...794..171T}
{The Polarbear Collaboration: P.~A.~R.~Ade} et~al., 2014, \mn@doi [\apj]
  {10.1088/0004-637X/794/2/171}, 794, 171

\bibitem[\protect\citeauthoryear{{Totsuji} \& {Kihara}}{{Totsuji} \&
  {Kihara}}{1969}]{1969PASJ...21..221T}
{Totsuji} H.,  {Kihara} T.,  1969, \pasj, 21, 221

\bibitem[\protect\citeauthoryear{{Wittman}, {Tyson}, {Kirkman}, {Dell'Antonio}
  \& {Bernstein}}{{Wittman} et~al.}{2000}]{2000Natur.405..143W}
{Wittman} D.~M.,  {Tyson} J.~A.,  {Kirkman} D.,  {Dell'Antonio} I.,
  {Bernstein} G.,  2000, \nat, 405, 143

\bibitem[\protect\citeauthoryear{{York} et~al.,}{{York}
  et~al.}{2000}]{2000AJ....120.1579Y}
{York} D.~G.,  et~al., 2000, \mn@doi [\aj] {10.1086/301513}, 120, 1579

\bibitem[\protect\citeauthoryear{{Zehavi} et~al.,}{{Zehavi}
  et~al.}{2002}]{2002ApJ...571..172Z}
{Zehavi} I.,  et~al., 2002, \mn@doi [\apj] {10.1086/339893}, 571, 172

\makeatother
\end{thebibliography}

\appendix

\section{Equivalence between mode projection and masking}
\label{sec:mp_mask}

Analyzing a specifically designed toy experiment, we now demonstrate
the conceptual equivalence of mode projection and masking. For a data
map with \npix pixels, where \npix is large, we consider a template
with real space representation $(\template)_i = \delta_{i j}$, i.e.,
only a single template pixel with index $j$ is different from zero. We
will now show that the mode projection algorithm yields identical
results compared to a PCL analysis, where pixel $j$ has been masked.

Applying the cleaning procedure \eq{eq:filtered_data} to obtain the
filtered data vector, we find
\equ{
  (\tdata)_i =
  \begin{cases}
    0 & i = j \\
    \data_i & \mathrm{otherwise.}
  \end{cases}
}
To debias mode projection results requires the exact knowledge of the
template power spectrum, a quantity that will depend on the details of
the pixelization scheme in our test case (e.g., pixel shape, size,
position, assumed sub-pixel model). In favour of a fully analytical
treatment of the problem, however, we choose to work with an
approximate expression of the template power spectrum instead,
\equ{
  \clt \approx \frac{4 \pi}{\npix^2} \, .
}
Since maximally localized fields in real space in general do not
possess a well-defined band limit in Fourier space, we further impose
a hard limit at \lmax, chosen such that the total number of Fourier
modes equals the number of pixels, $(\lmax + 1)^2 = \npix$, finding
\equ{
  \sum_{\ell} \llpo \clt = \frac{4 \pi}{\npix} \, .
}
The matrix used to debias power spectrum measurements with mode
projection (\eqs{eq:debiasing_matrix}{and}{eq:bias_matrix}) then takes
the simple form
\equ[eq:mp_pcl_equivalence_1]{
  (\one + B)_{\ell_1 \ell_2} = \left( 1 - \frac{2}{\npix} \right) \cdot
  \delta_{\ell_1 \ell_2} + \frac{2 \ell_2 + 1}{\npix^2} \, .
}

After a full analysis of the template projection algorithm for this
specific case, we now derive the corresponding equations for a PCL
power spectrum estimator. Masking the input map will set the pixel
with index $j$ to zero while leaving all other entries
untouched. Using identical assumptions as before, the power spectrum
of the mask $W = 1 - \template$ is approximately given by
\equ{
  \clh{\mask} \approx
  \begin{cases}
    4\pi \left( 1 - \frac{1}{\npix} \right)^2 & \ell = 0 \\
    \frac{4 \pi}{\npix^2} & \mathrm{otherwise.}
  \end{cases}
}
Using this expression, we obtain for the PCL coupling matrix,
see \eq{eq:coupling_matrix} below,
\eqa[eq:mp_pcl_equivalence_2]{
  M_{\ell_{1} \ell_{2}} &= \frac{2\ell_{2} + 1}{4 \pi} \left[
  4\pi \left( 1 - \frac{2}{\npix} \right)
  \wigner{\ell_{1}}{\ell_{2}}{0}{0}{0}{0}^{2} \right. \nn
  &+ \left. \frac{4 \pi}{\npix^2} \sum_{\ell_{3}} \llpo[_{3}]
  \wigner{\ell_{1}}{\ell_{2}}{\ell_{3}}{0}{0}{0}^{2} \right] \nn
  &= \left(1 - \frac{2}{\npix} \right) \cdot \delta_{\ell_{1}
    \ell_{2}} + \frac{2 \ell_2 + 1}{\npix^2} \, .
}

Finding identical results for the filtered (mode projection) or masked
(PCL power spectrum estimation) data vector as well as for the
debiasing procedures
(\eqs{eq:mp_pcl_equivalence_1}{and}{eq:mp_pcl_equivalence_2}), we
conclude the full equivalence of the two schemes.

We note in closing that the effect of any binary mask can therefore be
interpreted as projecting a collection of template maps. In this case,
each template would be non-zero only for a single pixel that falls
inside the masked area. For more general weight maps that are not
restricted to the numerical values zero and one, this equivalence is
no longer true and we have to resort to the more complicated schemes
discussed in \sects{sec:sub_cut_sky_single} and
\ref{sec:sub_cut_sky_multi}.

\section{PCL coupling kernels}
\label{sec:app_pcl}

We start from the Gaunt integral that allows to express the product of
three spin-0 spherical harmonics in terms of Wigner 3j symbols
\citep{Gaunt151},
\eqa[eq:gaunt]{
  \int \dn \, Y_{\ell_1 m_1}(\bn) &\, Y_{\ell_2 m_2}(\bn) \, Y_{\ell_3
    m_3}(\bn) \nn
  &= \left[ \frac{\llpo[_1] \llpo[_2] \llpo[_3]}{4 \pi} \right]^{1/2} \nn
  &\times \wigner{\ell_{1}}{\ell_{2}}{\ell_{3}}{0}{0}{0}
  \wigner{\ell_{1}}{\ell_{2}}{\ell_{3}}{m_{1}}{m_{2}}{m_{3}} \, .
}
Given this useful relation, the following Fourier space representation
of the coupling kernel can straightforwardly be obtained from
\eq{eq:pcl_mode_coupling} \citep{2002ApJ...567....2H},
\eqm[eq:coupling_kernels]{
  K_{\ell_1 m_1 \ell_2 m_2 } \\ = \sum_{\ell_3 m_3}
  \alm[_3][_3][\mask](-1)^{m_2} \left[ \frac{(2\ell_1 + 1)(2\ell_2 +
      1)(2\ell_3 + 1)}{4 \pi} \right]^{1/2} \\
  \times \wigner{\ell_{1}}{\ell_{2}}{\ell_{3}}{0}{0}{0}
  \wigner{\ell_{1}}{\ell_{2}}{\ell_{3}}{m_{1}}{-m_{2}}{m_{3}} \, ,
}
where we have introduced the spherical harmonic coefficients of the
mask, $\alm[][][\mask]$.

Making use of the orthogonality relations of the Wigner 3j symbols
\citep[e.g.,][]{edmonds1996angular}, the coupling matrix that connects
the ensemble average of full- and cut-sky power spectra is given by
\equ[eq:coupling_matrix]{
  M_{\ell_{1} \ell_{2}} = \frac{2\ell_{2} + 1}{4 \pi} \sum_{\ell_{3}}
  \llpo[_{3}] \clh[_{3}]{\mask}
  \wigner{\ell_{1}}{\ell_{2}}{\ell_{3}}{0}{0}{0}^{2} \, .
}
It is a function of the mask power spectrum $\clh{\mask}$ only.

\bsp
\label{lastpage}
\end{document}